\documentstyle[aps]{revtex}

\begin{document}

\title{Linear and Nonlinear Theory of Eigenfunction Scars}
\author{L. Kaplan\thanks{kaplan@physics.harvard.edu}
 \\Department of Physics and Society of Fellows,\\ Harvard
University, Cambridge, MA 02138\\ \vskip 0.1in
E. J. Heller\thanks{heller@physics.harvard.edu}
 \\ Department of
Physics and Harvard-Smithsonian Center for\\
Astrophysics, Harvard University, Cambridge, MA 02138}
\maketitle

\begin{abstract}  The theory of  scarring of eigenfunctions of classically
chaotic systems by short periodic orbits is extended in several ways.
The influence of short-time linear recurrences on
correlations and fluctuations at long times is emphasized.
We include the contribution to scarring of
nonlinear recurrences associated with homoclinic
orbits, and treat the different scenarios of random and nonrandom long-time
recurrences.
The importance
of the local classical structure around the periodic orbit is emphasized,
and it is shown for an optimal choice of test basis in phase space, scars
must persist in the semiclassical limit.
The crucial role of
symmetry is also discussed, which together with the
nonlinear recurrences gives a much  improved account of the actual strength
of scars for given classical orbits and
in individual  wavefunctions. Quantitative measures of scarring
are provided and comparisons are made with numerical data.

\end{abstract}

\section{Introduction}

Some years ago Berry\cite{berry1} suggested that quantum  eigenstates of
classically
chaotic systems should locally look like
random superpositions of plane waves of the same (local) wavevector
magnitude  $k$, producing Gaussian random fluctuations in position space.
This is the natural quantum manifestation of complete classical
uniformity on the  energy hypersurface for chaotic systems. It is also a
consequence
of random matrix theory (RMT), assuming that this theory
applies to classically chaotic
systems\cite{bohigas}.  On the other
hand, Gutzwiller periodic orbit theory of the energy spectra of classically
chaotic
systems  has enjoyed much success, and it would  be strange if periodic
orbits had
no visible manifestation in the eigenstates.   Indeed, exact solutions of the
Schr\"odinger equation for classically chaotic systems  sometimes exhibit
obvious and
apparently nonrandom patterns of concentration on periodic orbits.
Other non-coordinate space representations such as projections  onto
coherent states
(Husimi phase space)  naturally show the same effect.  This is a fully quantum
 phenomenon, although the explanation of scarring in terms of classical
periodic
orbits and their stability exponents\cite{scar1} is based
on semiclassical arguments.  It is known that the
structure of individual quantum eigenstates can often be well reproduced
in the semiclassical approximation, ignoring ``hard quantum" effects
such as tunneling and diffraction\cite{tomsovic}. Scars are
surprising because  classically (in the absence of phase coherence)
there is no such accumulation of density
near a periodic
orbit in the long time limit for a bounded chaotic system.  Scars are
interesting
because they  stand out against the monotonous  backdrop of wavefunction
randomness over most of phase space.
It has been shown that explicitly constructed ``random''
wavefunctions (random superpositions of plane waves
with constant wavevector magnitude) show a
  ridge pattern that can be thought of as a precursor of true
scarring\cite{scarlet}.

Scars have now been seen in many studies, including experimental work
in microwave cavities\cite{sridhar,stockman}, tunnel junctions\cite{fromhold},
and the hydrogen atom in a uniform magnetic field\cite{wintgen,wintscar}.

To the casual observer the scarring phenomenon finds its proof in
visual evidence ({\it e.g.} heavily scarred wavefunctions when
plotted in coordinate
space).
However there are quantitative reasons for scarring; the first theory
for them\cite{scar1}  was based on wavepacket motion (in the semiclassical
limit)
along a periodic orbit, and used the {\it linearized} short time  dynamics
near the
orbit. This was a Husimi  phase space theory; subsequent work by
Bogomolny\cite{bogo}
and Berry\cite{berry2} were coordinate space and Wigner phase space theories,
respectively.  These also were based purely on the linearized dynamics
in the vicinity of an unstable periodic orbit.
 Ref. \cite{scar1} showed that
scar strength for the heavily scarred states,
as measured by the ratio of actual density to statistically
expected
density, was expected to be  a function of the Lyapunov exponent
$\lambda$ only, tending to
$c/\lambda$ for small $\lambda$, where $c$ is a constant.
(Note the independence of $\hbar$.)
That paper also
remarked that scars are sometimes much stronger than this.  At the other
extreme,
Steiner's work with  systems of constant negative curvature was claimed  to
show
no scarring
when it  was supposed to exist\cite{steiner}.  These factors plus confusion
over the
definition and measures of scar strength have understandably caused much
discussion
of whether there is indeed a theory of scar strength.  The work by Fishman,
Agam,
and co-workers\cite{agambrenner,agamfishman,fishprange} has provided additional
perspectives, and a proposed measure for scars, about which
we comment later.

The main points  of this paper are (1) to point out that very often the
original linear theory, together with a proper account of gaussian
fluctuations and
symmetry is sufficient to understand scar strength;  (2)  to show that
scarring   stronger  than this may  sometimes be
understood in terms of identifiable  {\it
nonlinear}
homoclinic recurrences  associated with a given periodic orbit, the effect
of which recurrences will turn out to be $\hbar$-dependent;
(3) to firm up the existing definitions and measures of scarring; and
(4) to extend the notion of scarring to  classical structures
associated with periodic orbits, as suggested already by the
work of Voros\cite{voros}.
 We do not
attempt here to treat the case of many periodic orbits of similar
stability, period, and action
contributing within one $\hbar$ cell in phase space. This can also
dramatically
enhance scarring, but it is a non-asymptotic effect which
disappears in the
$\hbar\to 0$ limit.

In the following section, we hope to clarify the concept of scarring.
Then we provide a review of the original linear theory,
 for completeness and to establish the important
concept of the spectral envelope, which we use heavily.
After presenting the properties of scars as a
localization phenomenon, we turn to
considerations of discrete symmetries,  and of nonlinear fluctuations about the
envelope,
which together explain many cases of enhanced wavefunction
scarring.  Connections with semiclassical theory are made here, and
emphasis is placed on the constraints which the short-time dynamics places
on the stationary properties of the system.
Various measures of scarring are described, including ones which
explicitly incorporate the linearized dynamics around a classical orbit.
This is followed by numerical studies which show the expected
amount of scarring
according to the linear theory supplemented by gaussian random fluctuations,
in the first statistical study of
a large number of scarred states.
Numerical evidence is also presented for the effects on scarring of
individual homoclinic orbits, in a situation where the long-time recurrences
are not random. We end with  a discussion of the
state of the  theory of wavefunction scarring.

\section{What is a scar?}

We begin with a list of things which are {\it not} properties of scars,
followed by a list of properties which are in fact associated with scars.
Several of these statements will gradually become clearer
in the theory sections
to follow.

\vskip 0.1in
\noindent{What scars are {\it not}:}
\begin{itemize}

\item { Scarring is not associated with stable classical orbits.  These
do of course attract wavefunction density strongly in some eigenstates, but
the
reasons are well understood in terms of the semiclassical theory of integrable
systems\cite{berrytabor}. The scarring phenomenon is also qualitatively
distinct from the ``bouncing ball" states associated with marginally
stable orbits\cite{bb}.}

\item { Weak scars are not always visible to the ``naked eye''; there is
scarring nonetheless according to the definition given below. The amount
of scarring associated with specific eigenstates varies significantly from
state to state, though in accordance with a theoretically predicted
distribution. Also orbits with larger instability exponents exhibit
less scarring on average, and a statistical analysis may be necessary
in such cases to determine that scarring is indeed present.}

\item { Scarring is not a threat to ergodicity of wavefunctions in the
sense of Schnirelman, Zelditch, and Colin de Verdiere\cite{SZCdV},
because the phase space area affected by scarring vanishes in the
semiclassical limit, forming a narrower and narrower region
around the periodic orbit. }

\item {Scars are not merely associated with a one dimensional line along a
periodic
orbit; rather, they are associated also with the stable and unstable manifolds
of that orbit. For this reason, a phase space study of scarring  may often be
more illuminating than a coordinate space projection.}

\item{ Scars do not disappear as $\hbar\to 0$, except in the sense that
the total amount of scarring is expected to become distributed over an ever
increasing
number of eigenstates in that limit, while the region of phase space
in which the eigenstates are scarred is simultaneously decreasing.}

\end{itemize}

\noindent{What scars are: }

We begin with a definition, close to what was already
given in the 1989 Les Houches proceedings\cite{leshouches}:
\begin{itemize}

\item{{\it Definition:} A quantum eigenstate of a classically chaotic
system has a {\it scar} of a periodic
orbit if its density on the classical invariant manifolds
near the periodic orbit  is enhanced over the statistically expected density.}

\item{Alternatively, an unstable periodic orbit is scarred when
some eigenstates of the system have greater amplitude, and others
less amplitude, along the orbit than would be predicted based on
gaussian random fluctuations. Also, a wavepacket launched on or near
such an orbit will have a tendency to return to the orbit, having
larger overlaps with itself at long times than
a wavepacket launched elsewhere in phase space.}

\item{Scars can appear as strong enhancements in the eigenfunction
coordinate space
density surrounding periodic orbits, especially near self-conjugate
points along the classical orbit, as shown by Bogomolny\cite{bogo}.}

\item{Scar strength $S$, as measured by the
projection of scarred eigenstates
onto a coherent state centered on the scarring periodic orbit
and aligned along the stable and unstable manifolds, is
generically a function only of $\lambda$, the Lyapunov exponent for one
period of the periodic orbit, and is $\hbar$-independent.
For small $\lambda$, $S\to C/\lambda$, where
$C$ is a constant obtained by considering the linear theory at short times
combined with random long-time fluctuations. }

\item{Enhancements of the scarring phenomenon
can occur in the presence of strong, isolated long-time
recurrences associated with homoclinic orbits. Symmetry factors
also must be included if one is to obtain a quantitatively correct
picture of scarring.}

\end{itemize}

\section{Linear theory of scarring}
\label{lin}
For completeness and context we need to
review the linear theory of scarring, first discussed in
Ref.~\cite{scar1}.
Consider an unstable fixed point of a classical map located at the origin,
with the stable and unstable manifolds oriented along the $p$ (vertical)
and $q$ (horizontal) axes, respectively. (We will adopt this choice
of coordinate system throughout. In general, a canonical transformation
needs to be performed, {\it e.g.} in the case of an inverted
harmonic oscillator, to rotate and skew the stable and unstable
manifolds into this alignment). Linearizing the map
around the fixed point, we obtain to first order
\begin{eqnarray}
q' &=&e^{\lambda t}q \nonumber \\ p'&=&e^{-\lambda t}p \,,
\label{clasdyn}
\end{eqnarray}
where $\lambda$ is the Lyapunov exponent for one iteration of the
orbit. (The situation is very similar in the case where the stable
and unstable manifolds are  not thus aligned,
and analogous
formulae can be derived for that scenario. Also, for simplicity we do not
discuss the case of a wavepacket centered near, but not on, a periodic
orbit
(see Ref.~\cite{scar1}).)
We now take a gaussian
wavepacket
\begin{equation}
\label{wavepkt}
g_\sigma(q)=({4\pi\hbar^2 \over \sigma^2})^{1/4} e^{-q^2/2\sigma^2} \,,
\end{equation}
which corresponds to a classical distribution centered on the origin
with width $\sigma$ in the $q$ direction and width
$\sigma_p=\hbar/\sigma$ in the $p$ direction
($\hbar \ll \sigma \ll 1$, {\it e.g.} $\sigma \sim \sqrt\hbar$).
Now for a small enough $\hbar$, the initial wavepacket and its short-time
iterates are contained within the linear regime, and we have
the time-evolved wavepacket
$g_t(q)= U^t g_\sigma(q)$ given by the expression above in Eq.~\ref{wavepkt}
with $\sigma$ replaced by $\sigma_t=e^{\lambda t} \sigma$ (here $U$
is a quantum operator corresponding locally to the classical
dynamics given by Eq.~\ref{clasdyn}). Classically
this corresponds to a horizontal stretching and vertical shrinking of
the gaussian distribution in phase space.

The overlap
\begin{equation}
A(t)=\langle g_t|g\rangle = {e^{i\theta t} \over \sqrt{\cosh(\lambda t)}}
\end{equation}
is easily found by Gaussian integration (notice that the autocorrelation
function $A(t)$ is independent
of $\sigma$, the width of the initial wavepacket). Here $\theta$
is the quantum phase associated with the fixed point (semiclassically
it is given by $S/\hbar$, $S$ being the action for one traversal of the
periodic orbit, plus Maslov phases arising from caustics).
This time domain behavior can be fourier transformed to obtain
an envelope in the (quasi-)energy spectrum, centered at $E=\theta$ and
with a width which depends only on  $\lambda$, scaling
linearly with $\lambda$ for small $\lambda$.

We remark here that the situation for a fixed point of period $P>1$ is similar.
In this case the linear autocorrelation function is nonzero only
at integer multiples of $P$, and the corresponding spectrum
has $P$ identical bumps, each of a width and height related to the instability
of the entire orbit (see also the discussion in Section~\ref{pgt1}).
Additional time scales are present in a continuous-time
system, which are not directly relevant to the phenomenon of scarring,
but which produce a background spectrum relative to which scarring
can manifest itself. These issues are addressed in Section~\ref{contitime}
of this paper.

In the case of exact linearity, or where the evolving wavefunctions are
allowed to escape to infinity at long times (as in an inverted
harmonic oscillator), the preceding is all that
there is to be
said about the spectrum of the wavepacket. The width of the
spectral bump then corresponds to a decay rate. But in a closed, unitary
system,
the escaping probability must eventually start returning to the origin. In
a classically mixing system, this will begin happening not later than by
the mixing time, this being
the time required for a classical distribution
corresponding to a minimum uncertainty wavepacket to spread through all
of phase space on a mesh of size $\hbar$. The mixing time scales as
$T_{\rm mix} \sim \log_{\overline\lambda}(N)$, where $\overline\lambda$
is the ``typical" exponent for the entire system, and $N$ is the total
number of states in the available phase space.

The key point is that the Fourier transform of $A(t)$
for small $\lambda$ localizes
the spectrum (local density of states) to a region of width $\sim
\lambda$, smaller than
the whole quasi-energy interval. In effect the initial state is
in  a resonance mode which decays more slowly than
a random state. A random state  should decay in a time of the order
of a single
time step for a discrete map. (The reason for the single step decay is
simple:
a random state having a random ({\it i.e.} RMT)
local density of states spectrum has a quasi-energy  uncertainty
of the whole interval ( $\delta \epsilon = \delta \omega/\hbar= 2\pi$).
Now
$\hbar \delta \epsilon \delta \tau \sim \delta \omega \delta \tau \sim \hbar$
implies  $\delta \tau \sim{\cal O} (1)$, {\it i.e.} one time step).
The corresponding resolved spectrum for a wavepacket launched on a
periodic orbit thus
{\it cannot } be picked from an {\it a priori} RMT local density of
states, as
Figure~\ref{envel} shows.  Now, to complete the point, we recall that
the  spectral line intensities are the squared projections of
eigenstates onto the local ``test'' Gaussian.  {\it The intensities
are the ``support'' of the envelope, and are thus required to
be (upon local average) larger in the peak region of the envelope than RMT
predicts
by  a factor of order $\lambda^{-1}$.} This enhancement of the
overlaps (that is, the enhancement by a factor of order $\lambda^{-1}$ over
the statistical expectation of $1/N$, where $N$ is the dimension
of the Hilbert space) means that at a minimum there must be states
with a projection onto the test state of order $\lambda^{-1}$ larger
than what is statistically expected.
However, if this projection were to be shared
in an egalitarian fashion among all the available states, then
most or all of the states in the peak region would be enhanced
by this factor.  On the other hand, if only a small fraction $f$
of the available states are enhanced, then these states must have
larger projections  ${\cal O}(\lambda^{{-1}} f^{-1})$ onto the test state
in order to support the local density of states envelope. These two
extremes are illustrated in Figure~\ref{envel2}.
The short time
dynamics, which depends only on the linear or ``tangent'' map
around the periodic orbit on which the test Gaussian
is centered, cannot tell us without further assumptions
which extreme (or intermediate)
regime is realized; it only tells us that some states must
be enhanced.  The egalitarian case corresponds to the least striking
type of scarring, since each state is enhanced at most by a factor
of order $\lambda^{-1}$.  If $\lambda$ is not too small, this enhancement
is not even competitive with the fluctuations expected from RMT, and
we might conclude by cursory
inspection that individual states are not scarred at all.
Indeed, this would be a justifiable definition, although the {\it
systematic}, statistically significant
enhancement of many nearby states in the egalitarian
case would still reveal the underlying mechanism of scar localization.
In effect this definition was adopted by Steiner and
co-workers\cite{steiner} in
their studies of the eigenstates of the hyperbolic billiards, which
appear to live close to the egalitarian limit.
In the opposite ``totalitarian'' extreme, enhancements are very large, and
scars are
obvious in pictures of eigenstates, even for orbits which are very unstable.
It is important to note however that for small   $\lambda$
 even the linear (short time)
theory in the ``worst case'' egalitarian scenario predicts strong
scarring, well above the typical RMT fluctuations, of strength $1/\lambda$.

These considerations extend easily to include the possible
dependence of scar strength on
$\hbar$ or on the
density of states.
If the density of states is such that
only one or a few states can exist within a quasi-energy  width $\lambda$,
then effectively only the ``totalitarian'' option exists.
This is a strong localization regime, where one or a few states carry
the total scar intensity.  At the ideal unitary limit of an overlap of $1$,
one
state is entirely localized to the periodic orbit region.
(This was the basis for our conclusion that the bouncing ball
modes in the stadium billiard persist up to infinite energy\cite{bb}).
Starting from this extreme, as $N$ increases, the scar strength of
individual eigenstates {\it
could} decrease as fast as $1/N$, in the  egalitarian limit, although
the intensity {\it enhancement factor} would still remain finite, as
the average intensity is also decreasing as $1/N$.  (In a billiard
system there is a $\sqrt{E}$ increase in the number of
affected states with increasing
$E$: the density of states is independent of $E$ but the
energy width $\delta E$ of the scar ``resonance'' scales as
$\sqrt{E} \lambda$, where
again $\lambda$ is the Lyapunov exponent for a complete period of the
orbit.  The time required to traverse this orbit goes as $1/\sqrt{E}$,
thus $\delta E \sim \sqrt{E} \lambda$).

We  remark that scarring can become no weaker than the egalitarian
limit defined above for any given periodic orbit, even as $\hbar \to 0$.
Suppose that the egalitarian
 limit is the usual circumstance as  $\hbar \to 0$.  Then scars become
 less dramatic but do not disappear as $\hbar \to 0$ as measured
 by the test states whose area in a surface of section is $h$.
 However this area (projected onto coordinate space, say) amounts to
 a diminishing portion (going as $\sqrt\hbar$ for a phase space Gaussian
with an aspect ratio of order unity) of the total
 coordinate space volume.  These subtleties have caused much confusion
 over whether scars ``disappear'' as $\hbar \to 0$.

\subsection{Husimi projections and phase space tubes}

The projection of  Gaussian wavepackets, or other distributions   localized
around periodic orbits, onto the eigenstates as a test of their
localization
properties was introduced in Ref.~\cite{scar1}.  Subsequently, the idea of
detecting and quantifying scars by integrating
over tubes in phase space  surrounding the periodic orbit has been
discussed\cite{agamfishman}.
 The tube should be of diameter
$\sqrt{\hbar}$ normal to the direction of the
orbit. The diameter originally used was $\hbar$ but
$\sqrt{\hbar}$ is more appropriate and is used
in more recent work (S. Fishman, private communication). Of course, the
structure of the linear local dynamics around the periodic orbit must
also be considered here. In particular, for certain alignments
of the stable and unstable manifolds with respect to
the $p$ and $q$ directions, a tube of width
$\hbar$ in position space and width $1$ in momentum would be equally
optimal. This is
consistent with the findings of Li~\cite{li}, where certain orbits
in a stadium billiard show
optimal
scarring in coordinate space with a tube size scaling as the wavelength
(instead of as the square root of the wavelength).

The phase space tube approach
is closely related to the Gaussian wavepacket projection, as
we shall now show.
In two dimensions,
suppose we average the Gaussian projections over the whole
length of the periodic orbit (instead of taking an overlap with a
Gaussian centered at just one periodic
point). The mean wavepacket momentum points
along the orbit.  Then we have, for an  orbit pointing along the
x-axis, the average projection $S$ given by
\begin{eqnarray}
S={1\over L_{x}}\int \ dx_{0}\vert \langle \alpha (x_{0},p_{x0},y_{0},0)\vert
\psi_{E}\rangle\vert^{2} &=& {\rm Tr} (\rho_{L} \vert
\psi_{E}\rangle\langle \psi_{E}\vert ) \,,
\end{eqnarray}
where
\begin{eqnarray}
\rho_{L}(x,y,x',y') &=& {1\over L_{x}}\int \ dx_{0}
\exp[-(x'-x_{0})^{2}/2\sigma_{x}^{2}\hbar
-(x-x_{0})^{2}/2\sigma_{x}^{2}\hbar \nonumber \\  &&
-(y'-y_{0})^{2}/2\sigma_{y}^{2}\hbar -(y-y_{0})^{2}/2\sigma_{y}^{2}\hbar+
 i(x-x')p_{0x}/\hbar ]\nonumber \\  &\sim&
 \exp[-(x'-x)^{2}/4\sigma_{x}^{2}\hbar
-(y'-y_{0})^{2}/2\sigma_{y}^{2}\hbar -(y-y_{0})^{2}/2\sigma_{y}^{2}\hbar+
 i(x-x')p_{0x}/\hbar ] \,.
 \end{eqnarray}
Now  we Wigner transform this density matrix:
\begin{eqnarray}
  \rho_{L}^W( {\bf q}, {\bf p} ) \ &=& 2^d
 \int\limits_{- \infty}^\infty e^{-2i {\bf p}\cdot{\bf s}/ \hbar}
 \rho_{L}({\bf q +s,q-s})\, {\rm d}  {\bf s} \nonumber \\  &=&
 \rho_{L}^W( x,p_{x};y,p_{y} )   \sim \exp\left
[{-(y-y_{0})^{2}/\sigma_{y}^{2 }\hbar-
\sigma_{y}^{2}(p_{y}-p_{y_{0}})^{2} /\hbar
-\sigma_{x}^{2}(p_{x}-p_{x_{0}})^{2}/\hbar
} \right ].
\end{eqnarray}
We see {\it this is a phase space tube surrounding the classical
trajectory, of diameter $\propto \sqrt{\hbar}$}, if $\sigma_{y}^2$, the
aspect ratio in the $y$--$p_y$ subspace, is chosen to be of order unity.
Hence, the relation between the Fishman {\it et al.} phase space tube
and the Husimi projection is that the tube is the Husimi projection
averaged over the entire length of the orbit.  The advantage of this smearing
is the same as its disadvantage: it is insensitive to the
local direction of the classical invariant manifolds.  In this
sense the tube  is a somewhat duller probe for scarring,
while at the same time it has the advantage of providing a more universal
description of scars.

\subsection{Limitations of the linear theory}

If the initial
periodic orbit is sufficiently unstable, a ``quiet time" regime may exist
after the linear autocorrelation function has decayed and before the nonlinear
recurrences have had a chance to build up. This effect is not included
in the linear theory and will be reflected in the
absence of features in the spectrum corresponding to these time scales.
(See also the discussion in section \ref{contitime}.)
In the semiclassical limit, however,
the quiet time is very short compared to the
Heisenberg time $T_H \sim N$, where individual eigenvalues begin to be
resolved, and where the quantum mechanics begins to be
quasiperiodic.  (At
time $T_H$ the information contained in the quantum mechanics
gets exhausted, and the autocorrelation function at longer times can
be reconstructed using only the information for times up to $T_H$.)
Nonlinear recurrences at times between the end of the quiet time regime
and the onset of Heisenberg-scale dynamics are crucial for a proper
understanding of scarring, and will be the subject of Section~\ref{nonlin}
of this paper.

This section has been a review of the ``existing''
theory with remarks aimed at more recent work.
There are shortcomings of this theory which we aim to
remedy below:

\begin{itemize}

\item The linear theory does not say how the localization,
predicted from the short time dynamics, actually manifests
itself in the long time behavior of the autocorrelation function $A(t)$.

\item The linear theory can have no information about whether the
totalitarian or egalitarian limit of scar intensity distribution over
the available eigenstates is approached.  This information comes
from longer time dynamics which necessarily involves more than
the linearized tangent map of the periodic orbit.

\item No systematic study of scar strength over a large enough
ensemble to unambiguously test the  predictions of the linear theory
has yet been undertaken.
\end{itemize}

\section{Scarring as a localization phenomenon}
\label{scarlocal}
It is well known that short time coherences in
amplitudes lead to weak localization effects.
An example is provided by coherent backscattering in random
media, where a factor of 2 enhancement is realized
from the fact that a quantum path and its time reversed
image contribute equally, with the same phase.  Ensemble
averaging cannot remove this effect.

Scarring also involves a weak
localization effect resulting from  short time correlations which
depend on the properties of classical  {\it unstable} periodic orbits, and also
symmetry,
 with  consequences similar to those of coherent backscattering.
The short time coherences have an effect on the long-time,
stationary properties.  In the time domain, this means that the return
probability at long times for an initial state launched on an unstable
periodic orbit is enhanced from what one would expect naively in the
absence of this classical information. In the energy domain, a wavepacket
located near a periodic orbit will have enhanced, non-random overlaps with the
available eigenstates of the system. These enhanced
overlaps imply that scarring exists, although there is a subtle issue
of how the ``scar strength'' is to be shared amongst the eligible
eigenstates.  This issue of scar sharing is key to understanding
the strength of scars in individual eigenstates.  We will describe
briefly the concept of weak localization from the time and energy  points
of view and the
connection between them (A fuller discussion can be found in Ref.
\cite{qerg}),
and then
go on to discuss how such an effect can arise from short-time classical
behavior.

Consider a compact classical phase space of area $A$,
with chaotic dynamics given by a discrete-time
evolution (area-preserving map of $A$ onto itself),
and no conserved quantities.
 A two-dimensional
billiard  can be reduced to such a discrete time one-dimensional map by the
surface of section technique.
(Although we restrict ourselves to one spatial
dimension for specificity, the concepts are completely
generalizable.)
Our results can
be extended to a situation in which conserved quantities (such as energy)
are present, by
considering flow between phase space-localized states (usually coherent
states), and taking account of the energy spread contained in such
states. This problem is treated in Ref.~\cite{ergodic}, and is also mentioned
briefly in Section\ref{contitime} of this paper. We will
assume in the present discussion
that all of phase space is classically accessible from any smooth
starting distribution.

If the area is an integer multiple of Planck's constant $h$, $A=Nh$,
the system can be quantized (with a choice of quantization conditions), to
obtain an $N$-dimensional Hilbert space. Because the underlying classical
dynamics is completely ergodic, one might expect the eigenstates to appear
random in any natural basis, such as that of position, momentum, or
Gaussian states. Thus, let $|a\rangle$ be such a physically-motivated basis
and $|n\rangle$ be the basis of eigenstates. Then we expect the overlaps
$f_{an}=\langle a|n \rangle$
to be Gaussian variables with the normalization
condition $<|f_{an}|^2>=1/N$.
This does
{\it not} mean that all energy eigenstates have equal overlaps with
all the trial basis states, {\it i.e.} $|f_{an}|^2\ne 1/N$ for all $a, n$.
In fact, such a Gaussian distribution (predicted by random
matrix theory, which is based on the absence of a preferred basis for
analyzing the dynamics), leads to $<|f_{an}|^4>=F/N^2$, where $F=3$
if both $|a\rangle$ and $|n\rangle$ are real
(convenient if, for example, the dynamics is
time-reversal invariant), and $F=2$ otherwise. This is a quantum fluctuation
result and is already a
deviation from the classical expectation  of $F=1$. Localization, however,
is taken to mean an additional deviation of the $f_{an}$ distribution,
away from Gaussian
form, towards a distribution with a longer tail.
In particular, the inverse participation value (IPR) $<N^2|f_{an}|^4>$
(where the average can be taken over trial states, energy eigenstates,
or both, and also over an ensemble of systems) is in the presence of such
localization enhanced from its
ergodic value of $F$. Higher moments and the behavior of the tail
can also be investigated.

An important point is that scars entail a
correlation between overlap probabilities $|f_{an}|^2$ and the energies
$E_n$, for a wavepacket $|a\rangle$ located on a periodic orbit (such
correlations are absent under the assumptions of RMT).
In fact, in the
case of random new long-time
recurrences superimposed on top of semiclassically understood
short-time dynamics, the distribution of $f_{an}$ overall and
as a function of $E_n$ can be predicted analytically, in terms of that
short-time dynamics, as will be shown
below in Section \ref{randfluct}.

Let us now consider the connection with the time domain. We define the
autocorrelation function $A(t)=\langle a|U^t|a\rangle$,
where $U$ is the discrete
time evolution operator (a completely analogous notation can be written
down for continuous time). The fourier transform of $A(t)$ is
the weighted spectrum (local density of states)
$S(E)=\sum_n \delta(E-E_n) |f_{an}|^2$.
The squared autocorrelation function
\begin{equation}
|A(t)|^2=\left|\sum_n |f_{an}|^2 e^{-iE_nt}\right|^2
\end{equation}
can be thought of as a wavepacket-specific form factor, similar to
the usual spectral form factor $F(t)=\sum_{mn} e^{-i(E_n-E_m)t}$, but weighting
each term by the heights of the corresponding lines in the
spectrum $S(E)$ above.

For long times, in the
absence of degeneracies, one easily obtains the relation
\begin{eqnarray}
\label{ipr}
<|A(t)|^2>_t & \equiv & \lim_{T_{\rm max} \to \infty} {1 \over T_{\rm max}}
\sum_{t=0}^{T_{\rm max}-1} |A(t)|^2 \nonumber \\ & = & \sum_n |f_{an}|^4 \,,
\end{eqnarray}
where on the left hand side a time average must be  taken over times
long compared to the Heisenberg time $T_H$ (generically $T_H \sim N$).
Within RMT, both sides of Eq.~\ref{ipr} are predicted to approach $F/N$,
in the semiclassical limit $N\rightarrow \infty$. Localization is associated
with an enhancement in the long-time return probability $<|A(t)|^2>_t$.
We will see in Section~\ref{nonlin} how this is possible in the case of
scarring. We will also see there that short-time unstable orbits induce
nontrivial correlations $<A^\ast(t+\Delta) A(t)>$ at long times $t$.
These will be seen to correspond to eigenvalue-eigenstate correlations
(through the formation of an envelope in the spectrum) in the energy domain.

\section{Beyond the linear theory}
\label{nonlin}

\subsection{Homoclinic orbits}
In keeping with our treatment of the linear theory of scarring, we will
now discuss long time recurrences in the autocorrelation function from
a semiclassical point of view.
Let us consider a homoclinic orbit which begins near the fixed point $(0,0)$
along the unstable manifold at large negative times and again approaches
the fixed point along the stable manifold at large positive times.
Specifically, let the orbit ${\cal HC}$
be given by $\{(q_t,p_t)\}_{t=-\infty\ldots\infty}$,
such that $(q_t,p_t)=
(a e^{\lambda t},0)$ for $t \to -\infty$ and $(q_{t'},p_{t'})=(0,
b e^{-\lambda t'})$ for $t' \to +\infty$. Then we claim
that a thin vertical strip cut out of
the initial Gaussian near $q=a e^{\lambda t}$ at time $t$ will intersect the
same Gaussian as a long horizontal strip at $p'=b e^{-\lambda t'}$ at a
much later time $t'$.
Note that because the wavepacket is contained well inside the linearizable
region around the fixed point, the dynamics from time $t$ to time $t'$ can
be divided into three parts. First, the tall, narrow distribution shrinks
vertically and stretches horizontally as  its center moves out horizontally
at an exponential rate along the unstable manifold
(for, let us say, $\tau_1$ steps). This is followed by
complicated nonlinear dynamics which eventually brings the center of
the distribution back into the linearizable region, this time along the
stable manifold of the fixed point. Now the part of the distribution which
is in the linear region again begins to stretch horizontally and shrink
vertically, becoming a narrow horizontal strip moving in towards the
original wavepacket. We will denote by $\tau_3$ the time spent in this
last stage of the evolution. The first and third parts of the dynamics allow
the breadth of the initial distribution centered on $(a e^{\lambda t},0)$
and the height of the final distribution centered on $(0,b e^{-\lambda t'})$
both to be small compared to the size of the Gaussian wavepacket.
All of this is illustrated in Figure~\ref{mani}.

The overlaps of the Gaussian with the vertical and horizontal strips, as
well as the effects of the linear dynamics in stages one and three are
easy to write down analytically. There is also an amplitude factor $Q$ which
measures the stretching of the distribution in the nonlinear stage of
the dynamics, from time $t+\tau_1$ until time $t'-\tau_3$. Finally, there
is a phase $\phi_{\rm nonlin}$ associated with this nonlinear excursion.
The total contribution to the wavepacket autocorrelation function at time
$t'-t$ coming from this homoclinic excursion is given by a product of five
factors:
\begin{equation}
A_{\cal HC}=
e^{-q_t^2/2\sigma^2}\cdot e^{i\tau_1\theta}e^{-\lambda\tau_1/2}\cdot
Q(t+\tau_1,t'-\tau_3)e^{i\phi_{\rm nonlin}}\cdot
e^{i\tau_3\theta}e^{-\lambda\tau_3/2} \cdot
e^{-p_{t'}^2/2\sigma_p^2} \,.
\end{equation}
The factors $e^{-\lambda\tau_1/2}$ and $e^{-\lambda\tau_3/2}$ are
instability factors associated with the linearized motion of the wavepacket,
while $e^{i\tau_1\theta}$ and $e^{i\tau_3\theta}$ are the corresponding phases.
The suppression factors $e^{-q_t^2/2\sigma^2}$ and $e^{-p_{t'}^2/2\sigma_p^2}$
are associated with the fact that the initial and final points of the
excursion are both off-center relative to the gaussian wavepacket.
The total correlation function $A(T)$ is given semiclassically by a sum
of terms of the form given above over all homoclinic excursions
of length $t'-t=T$.
\begin{equation}
\label{hcsum}
A_{\rm SC}(T)=\sum_{\cal HC} \delta_{t'_{\cal HC}-t_{\cal HC}-T} A_{\cal
HC} \,.
\end{equation}

\subsection{Effect of short-time dynamics}
\label{steffect}

One might think naively that the various contributions to the sum in
Eq.~\ref{hcsum} are all independent
and uncorrelated at long times, but this is not the case. In fact,
correlations are present both among the different contributions to the
autocorrelation function at a fixed time, and also among homoclinic
contributions of different excursion lengths $T$.
Let us consider the homoclinic
orbit of the previous subsection, with the mapping taking us from
$(a e^{\lambda t},0)$ to $(0,b e^{-\lambda t'})$  in the time interval
$T=t'-t$. Now of course the same homoclinic orbit takes the $\Delta_1$-step
iterate of the original point, $(a e^{\lambda (t+\Delta_1)},0)$ to the
$\Delta_3$-step iterate of the final point, $(0,b e^{-\lambda (t'+\Delta_3)})$,
in a time $T+\Delta_3-\Delta_1$. In particular, taking $\Delta_1=\Delta_3$,
we have a family of excursions, all of the same length, associated with
one homoclinic orbit. An important thing to notice is that all these
contributions come with the same phase, the extra phase in the final $\Delta_3$
steps approaching the fixed point being exactly canceled by the missing
$\Delta_1$ steps at the beginning of the trip (there is only one phase
associated with the fixed point, and it is the same in stages one and three).
(``1''  and ``5'' in Figure~\ref{mani} have
exactly the same phase relation as ``2'' and ``6''.)
The phase $\phi_{\rm nonlin}$ coming from the nonlinear dynamics in stage
two is, of course, independent of $\Delta_{1,3}$. Naturally, the exponential
prefactor
\begin{eqnarray}
& & e^{-q_{t+\Delta_1}^2/2\sigma^2}e^{-p_{t'+\Delta_3}^2/2\sigma_p^2}
e^{-\lambda(\tau_1-\Delta_1)/2}e^{-\lambda(\tau_3+\Delta_3)/2}
\nonumber \\
& &=e^{-(q_te^{\Delta_1\lambda})^2/2\sigma^2}
e^{-(p_{t'}e^{-\Delta_3\lambda})^2/2\sigma_p^2}
e^{-\lambda(\tau_1-\Delta_1)/2}e^{-\lambda(\tau_3+\Delta_3)/2}
\end{eqnarray}
will depend on $\Delta_{1,3}$, so only a finite number of the infinite
family of contributions for $\Delta_1=\Delta_3$ will be of significant
size (this number, the number of iterations for which one tends to stay
near the periodic orbit, scales as $1/\lambda$.)
However, all of these will add exactly in phase.  This is an
important difference between the classical and semiclassical long-time
dynamics of the
system. The presence of this coherence is the  underlying
 reason for the fact that in quantum mechanics
the probability for coming back to an unstable periodic orbit at long
times is enhanced over the classical value of $1/N$, which fact leads
to scarring in the eigenstate domain, as seen in Section~\ref{scarlocal}.

We also notice that if we take $\Delta_1\ne\Delta_3$, so the new excursion
length is different from the original one, the difference in phase
is given by $e^{i\theta(\Delta_3-\Delta_1)}$, and each excursion of
length $T$ also contributes to the autocorrelation function at
$T+\Delta\equiv T+\Delta_3-\Delta_1$,
with this extra phase and a somewhat different
amplitude prefactor. Thus, as will be explained in more detail in
the following section,  nontrivial correlations will be present in the
autocorrelation function at nearby times, $<A^{\ast}(T+\Delta)A(T)>$,
for small $\Delta$. This is an effect that is qualitatively easy to understand
even in a classical picture in terms
of a ``reloading'' of the original wavepacket. Because the original wavepacket
is centered on a fixed point of the map, any significant recurrence
at time $T$ is expected to be accompanied by a recurrence for all times
$T+\Delta$, where $\Delta$ is within the decay time associated with the
fixed point. In effect, any new long-time recurrences get convoluted
with the (linear) short-time dynamics of the system around the fixed point.
In the energy domain, this corresponds (by fourier transform) to a
{\it multiplication} of the original linear envelope by an oscillating function
associated with the long-time recurrence. In the following section,
these statements will be made more explicit and quantitative under the
assumption of randomness in the homoclinic orbits which lead to long-time
recurrences.

\section{Random recurrence model}
\label{randfluct}
\subsection{Correlations among homoclinic excursions}
In a chaotic system, the number of homoclinic orbits increases
exponentially with the excursion length at long times. Let us
for the moment assume that the nonlinear phases associated with the terms
in Eq.~\ref{hcsum} are uncorrelated at long times. We also assume
a uniform distribution of homoclinic points $q_t$ and $p_{t'}$ along the
unstable and stable manifolds. We then obtain for the average square of the
long-time semiclassical autocorrelation function
\begin{equation}
<|A_{\rm SC}|^2>_{\rm diag}={\cal N}\int dq_t \int dp_{t'} e^{-q_t^2/\sigma^2}
e^{-p_{t'}^2/\sigma_p^2} \,,
\end{equation}
where the ``diag" subscript indicates that we are working in
a diagonal approximation (no correlations between the homoclinic orbits).
The mean squared amplitude factor associated with the excursions as
well as the densities of homoclinic points along the two manifolds
have been incorporated into the normalization factor $\cal N$. This
normalization factor can easily be fixed by noticing that classically,
in the absence of any coherence effects, the return probability must
approach $1/N$ at long times.

We now correct the assumption made in the previous paragraph and include
the fact that, as discussed in the preceding section, the contributions
from homoclinic excursions associated with a single homoclinic orbit
all add in phase. We then have
\begin{equation}
<|A_{\rm SC}|>^2={\cal N}'\int dq_t \int dp_{t'} \left|\sum_{\Delta_1}
e^{-(q_te^{\Delta_1\lambda})^2/2\sigma^2}
e^{-(p_{t'}e^{-\Delta_1\lambda})^2/2\sigma^2}\right|^2 \,,
\end{equation}
where the normalization factor $\cal N'$ is given by
\begin{equation}
<|A_{\rm SC}|>^2_{\rm diag}={\cal N}'\int dq_t \int dp_{t'} \sum_{\Delta_1}
\left|e^{-(q_te^{\Delta_1\lambda})^2/2\sigma^2}
e^{-(p_{t'}e^{-\Delta_1\lambda})^2/2\sigma^2}\right|^2 =1/N \,.
\end{equation}
We now see that the enhancement factor $<|A_{\rm SC}|>^2/
<|A_{\rm SC}|>^2_{\rm diag}$ is given by
\begin{eqnarray}
\label{delta}
{\sum_{\Delta_1,\tilde\Delta_1} \int dq_t \int dp_{t'}
g_{\tilde\Delta_1}(q_t,p_{t'})g_{\Delta_1}(q_t,p_{t'}) \over
\sum_{\tilde\Delta_1} \int dq_t \int dp_{t'}
g_{\tilde\Delta_1}(q_t,p_{t'})g_{\tilde\Delta_1}(q_t,p_{t'}) } \nonumber \\
={ \sum_{\Delta_1} \int dq_t \int dp_{t'}
g_{\tilde\Delta_1}(q_t,p_{t'})g_0(q_t,p_{t'}) \over
\int dq_t \int dp_{t'} g_0(q_t,p_{t'})g_0(q_t,p_{t'})}\,,
\end{eqnarray}
where $g_0$ is a Gaussian distribution of width $\sigma$ in position
and $\sigma_p$ in momentum, and $g_{\Delta_1}$ is the same
distribution stretched by a factor of $e^{\Delta_1\lambda}$ horizontally
and compressed vertically by the same factor.  In Eq. \ref{delta}
we have used the relations
\begin{equation}
 \int dq_t \int dp_{t'}
g_{\tilde\Delta_1}(q_t,p_{t'})g_{\tilde\Delta_1}(q_t,p_{t'})=
\int dq_t \int dp_{t'}
g_{0}(q_t,p_{t'})g_{0}(q_t,p_{t'})
\end{equation}
and
\begin{equation}
\int dq_t \int dp_{t'}
g_{\tilde\Delta_1}(q_t,p_{t'})g_{\Delta_1}(q_t,p_{t'}) =
\int dq_t \int dp_{t'}
g_{\tilde\Delta_1-\Delta_1}(q_t,p_{t'})g_{0}(q_t,p_{t'}) \,.
\end{equation}
Returning  to the discussion following Eq.~\ref{wavepkt}
we notice that the overlap of the original Gaussian wavepacket $a(q)$
with the same wavepacket stretched by a factor $e^{\Delta_1\lambda}$
is just the linearized autocorrelation function
$A_{\rm lin}(\Delta_1)$. The overlap of the
corresponding classical distributions is given by $|A_{\rm lin}(\Delta_1)|^2$,
and finally we have
\begin{equation}
\label{enhanced}
<|A_{\rm SC}|>^2_{\rm coherent}= {1 \over N}\sum_{\Delta_1}
|A_{\rm lin}(\Delta_1)|^2 \,.
\end{equation}

A similar analysis can be performed for the case $\Delta\equiv
\Delta_3-\Delta_1\ne 0$, and there we find
\begin{equation}
\label{enhanced2}
<A^{\ast}(T+\Delta)A(T)>_{\rm coherent}={1\over N}\sum_{\Delta_1}
A_{\rm lin}^{\ast}(\Delta+\Delta_1)A_{\rm lin}(\Delta_1) \,.
\label{timecorr}
\end{equation}
This shows the unmistakable effect of the short-time correlations at long
times.
The expressions obtained here quantify the
connection between short-time and long-time behavior
which was already suggested in the previous section.

\subsection{Reloading: another point of view}

We can also understand the results of Eqs.~\ref{enhanced},\ref{enhanced2} in a
somewhat different way, as suggested by the ``reloading" picture of
Section~\ref{steffect}. We write $A(T)$ at long times as
\begin{eqnarray}
A(T)&=&A_{\rm new}(T)+A_{\rm reloaded}(T) \nonumber \\
&=& A_{\rm new}(T) + \sum_{\Delta\ne 0} A_{\rm new}(T-\Delta)A_{\rm lin}
(\Delta) \,,
\end{eqnarray}
where $A_{\rm new}(T)$ are random, uncorrelated Gaussian variables
with variance $1/N$, {\it i.e.} $<A^{\ast}_{\rm new}(T')A_{\rm new}(T)>=
\delta_{TT'}{1 \over N}$ in an ensemble average.
Then one easily obtains the results of Eqs.~\ref
{enhanced},\ref{enhanced2}. The assumption being made here is
that the new amplitude
coming back to the fixed point at long times
fills the original Gaussian evenly in an
unbiased way, so that the evolution of this newly returned amplitude is
equivalent to the evolution of the original Gaussian. We saw above that
this is rigorously true semiclassically for uncorrelated homoclinic orbits.
However, the picture presented here may apply also to situations where
large short-time recurrences arise for reasons other than periodic orbits.
These recurrences may be due to ``almost periodic orbits", {\it i.e.}
classical orbits that return to within a minimum uncertainty wavepacket
of the starting point in phase space. Localization effects due to short-time
recurrences arising from diffractive paths (``diffractive scarring")
may also be understandable in this general framework.

It is important to point out that the reloading idea is critically
dependent on the hyperbolic nature of the dynamics. With a bit of coarse
graining
hyperbolicity quickly erases certain information about location in phase space.
More specifically, the attraction of orbits to the unstable manifolds
means that location along the  contracting directions is relatively
unimportant.

\subsection{Effect on spectra}
\label{effspectr}

We now claim that the time domain correlations given in Eq.~\ref{timecorr}
are just the right correlations to produce spectral fluctuations which
multiply the original linear envelope.
(This is important because one might have thought that the oscillations
in the spectrum caused by long-time recurrences simply get added to
the original linear envelope, instead of multiplying it. This would
produce completely different spectral behavior on small energy scales.)
Indeed suppose the spectrum has the form
\begin{equation}
S(E)=S_{\rm lin}(E)S_{\rm fluct}(E) \,,
\end{equation}
where $S_{\rm lin}(E)=2\pi\sum_T e^{iET}A_{\rm lin}(T)$ is the spectrum
given by the linear dynamics, and $S_{\rm fluct}(E)$ is the fluctuating
part, with the property that $K\equiv<|S_{\rm fluct}(E)|^2>$ is an
$E-$independent constant (an ensemble average is implied in the definition
of $K$). Now we have
\begin{equation}
<A^{\ast}(t+\Delta)A(t)>=<\int dE' S^{\ast}_{\rm lin}(E')
S^{\ast}_{\rm fluct}(E')e^{iE'(t+\Delta)} \int dE S_{\rm lin}(E)
S_{\rm fluct}(E)e^{-iEt}> \,.
\end{equation}
Averaging over $t$, and inserting the expression for $S_{\rm lin}(E)$
from above,
\begin{eqnarray}
<A^{\ast}(t+\Delta)A(t)> & = & <2\pi \int dE |S_{\rm lin}(E)|^2
|S_{\rm fluct}(E)|^2 e^{iE\Delta}> \nonumber \\ & = &
<{1\over 2\pi} \int dE  |S_{\rm fluct}(E)|^2 e^{iE\Delta}\sum_{TT'}
e^{iE(T'-T)}A_{\rm lin}^{\ast}(T')A_{\rm lin}(T)> \,.
\end{eqnarray}
Finally, inserting the fluctuation intensity $K$ and performing the
energy integral, we obtain
\begin{equation}
<A^{\ast}(t+\Delta)A(t)> = K\sum_T A_{\rm lin}^{\ast}(T+\Delta)A_{\rm lin}(T)
\,,
\end{equation}
which agrees with the result in Eq.~\ref{timecorr} above, obtained
by taking into account the phase coherence of all the homoclinic excursions
associated with a single homoclinic orbit.

\subsection{Heisenberg time dynamics and symmetries}
\label{heissym}

Up until now we have been assuming that the nonlinear recurrences associated
with the homoclinic orbits are sums of random contributions, under
the constraint of correlations due to the short-time linear behavior
near the periodic orbit. The picture here is that random ``new" recurrences
get smeared out by a function associated with the reloading effect. This
leads to Gaussian random fluctuations in the spectrum at all scales short
compared to the inverse of the mixing time, all of these uncorrelated
fluctuations multiplying the initial linear envelope. But now we have
to include an additional constraint coming from Heisenberg time dynamics,
namely unitarity and discreteness of the spectrum. In the absence of scarring,
these long-time correlations cause the mean return probability $|A_{\rm QM}
(t)|^2$ to converge to a value of $2/N$ at long times, $3/N$ if the
eigenstates and the initial wavepacket are both purely real. We may
reasonably suppose that this long-time constraint, associated with the
statistics of very long excursions away from the periodic orbit, is independent
of the short-time behavior near the periodic orbit. Thus, if we suppose
that what we in the previous section called stage two of the dynamics does
not know about the trajectory's approach to the periodic orbit at times
$|t|\to\infty$, then the same long-time constraints should be present in
the case of scarring. So we finally obtain a discrete spectrum, with
intensities given by a $\chi^2$ variable of 2 degrees of freedom (1 degree
of freedom for purely real $f_{an}$), except that in the case of scarring
this RMT line spectrum multiplies the original short-time envelope. The
line height associated with energy $E_n$ is given by
\begin{equation}
|f_{an}|^2 = {1\over N} S_{\rm lin}(E_n)|r_n^2| \,,
\end{equation}
where $r_n$ is a Gaussian variable, real or complex, with variance one.
In particular, the first two moments of this distribution are given by
\begin{equation}
<|f_{an}|^2>= {1\over N} S_{\rm lin}(E_n) \;\; <|f_{an}|^4>={F\over N^2}
S^2_{\rm lin}(E_n) \,,
\end{equation}
where $F=2$ or $3$ as explained above. We may note here that the linear
spectrum
$S_{\rm lin}(E)$ is purely real because of the unitarity of the time evolution,
$A(-t)=A^{\ast}(t)$. This holds also for the full spectrum $S(E)$, and for
any smoothed spectrum obtained by fourier transforming $A(t)$ for all
times $|t|<T_{\rm max}$.

The inverse participation ratio for the wavepacket is given by
\begin{eqnarray}
\label{iprval}
N<\sum_n |f_{an}|^4>  &= &F \int dE \ S^2_{\rm lin}(E) \nonumber \\
& =& F \sum_T |A_{\rm lin}(T)|^2 \nonumber \\
& =& F \sum_T [\cosh(\lambda T)]^{-1} \\
& \to & cF/\lambda \;\; {\rm for \;\; small} \;\; \lambda \nonumber \,.
\end{eqnarray}
Note that the IPR enhancement factor is a number that depends only on the
Lyapunov exponent of the periodic orbit. Higher moments of the $f_{an}$
distribution can be computed easily in terms of $A_{\rm lin}(T)$, always
taking into account the proper quantum fluctuation factors.

If the spectrum $S(E)$ is ensemble-averaged while preserving the periodic
orbit ({\it i.e.} if we average over the nonlinear dynamics only), we
recover the original linear envelope spectrum, the same spectrum which
is present in the case of an open system.

Finally we need to mention here an issue that has been the cause of some
misunderstanding in the literature, namely the issue of spatial symmetries.
In symmetric systems like the stadium billiard, many periodic orbits
either remain invariant under a reflection operation, or they get shifted
by some fraction of the total period,
or else they get mapped to their time-reversed counterparts.
In all these cases, one expects extra correlations in the contributions
coming from the homoclinic excursions which are related by such a symmetry
operation. Thus, consider the simplest case of a parity symmetry, where
the initial wavepacket centered on the periodic orbit remains unchanged
under the symmetry. Of course we know from quantum mechanics that the
wavepacket can overlap only with states in the even part of the Hilbert
space, so the IPR is increased by a factor of two from the value that
would otherwise be expected. We can also see this effect semiclassically,
because each homoclinic excursion away from the invariant periodic orbit
has a counterpart related by parity, and the contributions from the
two have the same amplitude and phase, so they add constructively. More
complicated situations can be treated similarly, and we will not
go into the details here.  In the example described later in the
paper, in Section~\ref{genbak},
we have intentionally desymmetrized the system of
interest to eliminate the enhanced scarring effects.

\section{Hierarchy of spectral envelopes}
In the previous section, we have constructed a model in which an initial
linear envelope is multiplied by random fluctuations associated with
scales between the mixing time and the Heisenberg time, producing a
discrete spectrum with heights that have a Porter-Thomas distribution
multiplying a smooth envelope. In this section, the picture will be
extended to incorporate other time scales that may be present in the system.

\subsection{Isolated recurrences}
\label{isolrec}
Suppose an isolated homoclinic orbit exists which leads to an anomalously
large return amplitude at some time $T_{\rm isol}$ which is large compared to
the decay time of the linear evolution. If $T_{\rm isol}$ is also small
compared to the time at which the returning orbits begin to proliferate
exponentially, we obtain a hierarchy of scales in the spectrum. The
initial linear envelope of width $2\pi/T_{\rm lin}$ is
multiplied by oscillations
at a scale of $2\pi/T_{\rm isol}$. These oscillations, although they are
associated with nonlinear dynamics and are therefore $\hbar$-dependent,
can nevertheless be computed semiclassically without much trouble. One can
also readily include recurrences associated with iterates of the
time-$T_{\rm isol}$ excursion.
This produces a modified envelope, which remains stable until the random
oscillations begin at a later time $T_{\rm rand}$ (see
Figure~\ref{nonlinfig}.)

Because a reloading effect
is associated with the isolated homoclinic recurrence as well as with the
linear dynamics, the new, modified envelope will
multiply the Gaussian random fluctuations arising from long-time
dynamics.
 The final spectrum then has structure on at
least three scales. If we take this discrete spectrum  and divide through
by the linear envelope, we will see fluctuations on top of oscillatory behavior
at scale $2\pi/T_{\rm isol}$. If we proceed further to divide out by these
identifiable oscillations, we get something that looks like
a normal Porter-Thomas spectrum,
with no energy-intensity correlations. An example of this situation is
described in Section~\ref{genbak}, with the numerical results appearing
towards the end of Section~\ref{genbakernum}.

At this point it may be useful to discuss briefly the relationship
between the quantum spectrum and the semiclassical approximation.
The initial linear envelope can of course always be obtained semiclassically.
The effect of the isolated recurrences can also be computed semiclassically
as long as these occur before the breakdown of the semiclassical
approximation. The random fluctuations can in principle be computed
semiclassically up until the breakdown time, which may be as large
as a finite fraction of the Heisenberg time. However, in practice
this computation may be difficult for large values of $N$,
because of the exponential
proliferation of homoclinic orbits. (This ``exponential wall"
may be overcome in some systems, making use of the Heisenberg
uncertainty principle and of decaying time correlations in the presence
of chaos\cite{expwall}.) Eventually, the semiclassical
approximation does break down,
and the precise locations and intensities of the spectral peaks may differ
somewhat between the full quantum mechanics and the semiclassical
approximation. There is some ambiguity in the definition of the
semiclassical spectrum, due to the non-unitarity of the semiclassical
propagator, which has eigenvalues that lie off the unit circle and
possesses distinct
left and right eigenstates. However, it seems possible to define
the spectrum in such a way that the spectral statistics have the
properties expected from the discussion above\cite{scspec}.
The {\it statistical}
properties of the quantum spectrum may be very similar to those
predicted by the semiclassical approximation, even in the absence of
detailed state-to-state agreement.

\subsection{Hamiltonian systems and ``quiet time"}
\label{contitime}

In a continuous-time conservative system,
like a two-dimensional billiard, another
relevant time scale is present in addition to those discussed previously.
This scale is $\delta E$, the energy width of the initial wavepacket.
Clearly, the wavepacket only has a chance to overlap with those eigenstates
which lie in the allowed energy range. This can be expressed in terms
of an envelope of width $\delta E$ in the spectrum that all subsequent
oscillations have to multiply. In the time domain, this corresponds to
the very short-time autocorrelation function, which measures the overlap
of the initial wavepacket with itself before it has had a chance to travel
a distance comparable with its width. This energy spread scales as
$\delta E \sim p\sigma_p = p\hbar/\sigma_q$. For a small wavepacket
$\sigma_q$, much smaller than the system size $L$, this scale is well separated
from the time of the shortest periodic orbit. This very short time
overlap is then followed by a relatively long ``quiet period"
in which $A(t)$ vanishes. If the original wavepacket lives on a short periodic
orbit, eventually a recurrence occurs corresponding to that orbit (and
producing an infinite series of bumps under the $\delta E$ envelope),
possibly followed by more quiet time and/or isolated homoclinic recurrences,
and eventually random recurrences begin which must be convoluted with
all that has come before. In the energy domain, this corresponds
to a series of nested envelopes. Of course, the quiet time mentioned above
is present even in the absence of scarring, and leads to quantitative
predictions about the absence of spectral fluctuations (vanishing of the
weighted form factor) on those scales.

\section{``Linear eigenstates" and other measures of scarring}

\subsection{Linear eigenstate scarring}
In the preceding sections we have focused on scarring as measured by
a Gaussian wavepacket centered on a periodic point. This has led to
criteria for scarring such as the distribution of the overlaps
of this wavepacket with
the eigenstates of the system, the inverse participation ratio,
eigenstate-eigenvalue correlations, wavepacket form factors, and the
distribution of long-time return probabilities. But we know from
previous analysis of the linear approximation that scarring
is associated not only with a periodic point but also with the
direction of the classical stable and unstable manifolds near that point.
It is reasonable to expect that the hyperbolic nature of the classical
dynamics should be seen in the quantum dynamics and in the eigenstates
as well. But the Gaussian wavepackets that we have been using so far carry no
information even about the linear dynamics around the periodic point.

This serves as motivation for replacing the elliptic wavepacket
$\phi_\sigma(q)$
of Eq.~\ref{wavepkt} with the coherent superposition of Gaussians
\begin{equation}
\phi(q)=Z\int d\beta \ g_{\sigma_{\beta}}(q)\ e^{-|\beta|/T_0} \,,
\end{equation}
where $\sigma_{\beta} = \sigma_0 e^{\lambda\beta}$ and
where $\beta$ is a time parameter, $T_0$ is a time cutoff, $Z$
is a normalization factor, and
$g_\sigma(q)$ is a Gaussian given by Eq.~\ref{wavepkt}. The state $\phi$
lives on the entire hyperbolic structure surrounding the stable
and unstable manifolds, and not only on the fixed point itself.
In the limit $T_0\to\infty$, $\phi$ is by construction
an invariant state of the linearized
dynamics around the fixed point. In this sense, $\phi$ knows about not only
the the location of the fixed point, but also about the linear dynamics
near that point. The extent to which $\phi$ differs from being an eigenstate
of the full evolution is a measure of the nonlinear dynamics (whereas
the extent to which the original Gaussian $g(q)$ differed from being an
eigenstate is primarily a measure of the linear instability around the
periodic point). The states $\phi$ are the
``linear eigenstates'' (they would be true eigenstates if the dynamics
were purely linear hyperbolic about the fixed point).
  We can perform a very similar kind of analysis with $\phi$
as we did with the plain Gaussians $g$, in particular looking at the
$\langle n|\phi\rangle$ distribution,
the inverse participation ratio, the autocorrelation function, and the
spectrum. Of course we must choose a value $T_0$ which is within
the linearizable regime of the motion in order to get a sensible result.
In this regime, the autocorrelation function for $|T|<T_0$ is given by
\begin{eqnarray}
A_\phi(T) & = & \langle \phi | U^T | \phi \rangle \nonumber \\
& = & e^{i\theta T}|Z|^2 \int d\tau {1\over \sqrt{\cosh(\lambda\tau)}}
\int d\beta \ e^{-|\beta|/T_0} \ e^{-|\beta+\tau-T|/T_0} \,.
\end{eqnarray}
We also want $T_0$ to be large compared to the decay time of the original
Gaussian wavepacket, if we are going to see the maximum enhancement of the
scarring effect. (The original scarring, as measured for example by the IPR,
scales as $T_{\rm decay} \sim 1/\lambda$ for small $\lambda$,
this being the time for which the
linear autocorrelation function is large. The amount of ``linear eigenstate"
scarring, as measured by overlaps with $\phi$ rather than $g$,
will scale as $T_0$, assuming
$T_0$ is large and within the linearizable regime.) So we require
\begin{equation}
1 < \lambda T_0 < \log N = \log {1\over 2\pi\hbar}\,,
\end{equation}
which criterion can be satisfied for small values of $\hbar$.
It is important to recognize that the
state $\phi$ is  nothing special as far as random eigenstates are
concerned, and has the same {\it a priori} spectral statistics as
 the states $g$.  From the point of view of the true eigenstates
 however, the states $\phi$ are  sharp tools which have
 very large overlaps with specific scarred states.  The
 predicted scar localization scales as $T_{0}$, which can be much
 stronger than the $1/\lambda$ effect detected by $g$.  The state
 $\phi$  is much better tuned to the  structure of eigenfunctions
 than is $g$.

The linear eigenstates idea can be extended to coherent superposition of
Gaussians
which live on the hyperbolic manifolds just off the periodic orbit.  A similar
analysis can be made, resulting in the construction of   states which
in the inverted oscillator analogy correspond to
wavefunctions just above or just below the energy
of the barrier.  We have not carried this out, but it seems likely
to be relevant to the hyperbolic states discussed by Nonnenmacher and
Voros\cite{voros}.

Linear eigenstate scarring can be viewed as one of a variety of approximation
schemes, where one compares the approximate eigenstates with the eigenstates
of the full system and looks for correlations between the two, which indicate
that the approximation has some validity. In this sense, the linearized
approach, being a log-time approximation, lies between ordinary scarring,
which approximates real dynamics
only on time scales of order $1/\lambda\approx O(1)$ and a full semiclassical
calculation, which can remain valid for times of the order of the
Heisenberg time. Unlike the full semiclassical approximation, it has
a simple geometrical meaning, being associated with the set of linearized
trajectories
that  come close to, but do not hit, the periodic orbit which is the
central object of our analysis.

\subsection{Correlation between periodic points on the same orbit}
\label{pgt1}

Given a periodic orbit of a discrete-time map with period $P>1$,
we expect scarred eigenstates to live along the entire orbit and not
only at one of the periodic points. In other words, nontrivial
correlations should exist in eigenstate densities measured at different
points along a single periodic orbit, so that the entire orbit and not only
a section of it is scarred.

More specifically, let us consider a
cross-correlation function $A_{ba}(T)=\langle b|U^T|a\rangle$ for
wavepackets $|a\rangle$ and $|b\rangle$ centered on two periodic points
$x_a$ and $x_b$.
By arguments analogous to those used in Section~\ref{nonlin}, we find that
any trajectory
taking a homoclinic point on the unstable manifold near $x_a$ to a homoclinic
point on the stable manifold near $x_b$ adds coherently with other
such trajectories time-shifted along the same homoclinic orbit. The
amplification factor resulting from this is again given by the short-time
linear return of wavepacket $|a\rangle$ to itself at time $t=nP$ (or
equivalently of wavepacket $|b\rangle$ to itself -- notice
that the short-time autocorrelation functions for the two wavepackets
are equal, both coming from the same total instability and phase as
accumulated over a traversal of the entire orbit). So we have for the
time-averaged probability
\begin{eqnarray}
\label{corres}
<|A_{ba}|^2>_T & = & \sum_n |\langle a|n\rangle|^2 |\langle b|n\rangle|^2 \\
& \to & {1\over N}\sum_t [\cosh(\lambda t)]^{-1} \,,
\end{eqnarray}
where $\lambda$ is the Lyapunov exponent for the period-$P$ orbit. This
is to be compared with Eq.~\ref{iprval} obtained previously.

We can also write
\begin{eqnarray}
<|A_{ba}|^2>_T & = & \sum_n |\langle a|n\rangle|^2 |\langle b|n\rangle|^2
\nonumber \\
& = & <A^{\ast}_{bb}(T)A_{aa}(T)>_T \,,
\end{eqnarray}
where $A_{aa}$ and $A_{bb}$ are the autocorrelation functions of the
two wavepackets. So the result of Eq.~\ref{corres} can also be thought
of as expressing the correlation between the two autocorrelation
functions at long times. This is not surprising, given a one-to-one
correspondence that can be set up between trajectories leaving and returning
to the vicinity of $x_a$ and those that start and end near $x_b$.

\section{Model system: generalized baker's maps}
\label{genbak}

As a testing ground for our predictions about nonlinear
scarring, we will use
the generalized baker's maps, a class of bernoulli systems which are a paradigm
of hard chaotic behavior. In addition to having no stable regions of
phase space, these systems have the property that the long-time
semiclassical dynamics and the semiclassical eigenstates can be
computed efficiently ({\it i.e.} in a time that scales as a power
law, rather than exponentially in $1/\hbar$)\cite{expwall}. This is useful for
studying a phenomenon such as scarring, where predictions about the system
are made based on our expectations about the statistical properties
of the semiclassical behavior. These predictions, obtained in the
previous sections, can then be independently compared with the exact
semiclassical and the full quantum results, allowing us to distinguish
errors in the statistical arguments from errors inherent in the
semiclassical approximation itself. As we will see in
Section~\ref{genbakernum}, semiclassically
computed measures of scarring agree only roughly with the results of a full
quantum computation for any particular periodic orbit in a given realization
of a chaotic quantum system. The fluctuations of the semiclassically
computed scarring strengths (when considering an ensemble of orbits
with a fixed Lyapunov exponent)
are however virtually identical (in mean and variance) to those of the quantum
scarring strengths. Any anomalous scarring for a given orbit must be
attributed to correlations in new long-time recurrences near the Heisenberg
time, such correlations being present in both the semiclassical and
full quantum computations.

We will discuss briefly the definition and properties of the
classical, semiclassical, and quantum generalized baker's map, referring
the reader to the literature on the subject for more details
\cite{baker,bakeroth}.
Classically, the map is defined as a map of the unit square onto itself,
where the square is initially cut up into $M$ vertical strips with
widths $w_m$ ($\sum_{m=0}^{M-1} w_m=1$) and height
$1$. (In the original baker's
map, $M=2$ and the two strips each have width $1/2$. This leads to a constant
Lyapunov exponent of $\log (2)$ everywhere in phase space. In the generalized
version of these maps, this restriction is lifted, allowing different
periodic orbits of the same length to have different instability factors.)
Each strip is stretched horizontally
and compressed vertically, in an area-preserving way, so that the width
becomes $1$ and the height becomes $w_m$. The deformed horizontal strips
are finally stacked on top of each other, in some order (conventionally,
the left-to-right order of the initial arrangement corresponds
to the bottom-to-top order of the final one). If we define $s_m=
\sum_{j<m} w_j$ to be the left edge of the $m$-th strip, we have
\begin{eqnarray}
x'&=&(x-s_m)/w_m \nonumber \\
p'&=&w_m p + s_m
\end{eqnarray}
for $x$ in the $m$-th strip, $s_m\leq x<s_{m+1}$. Trajectories and
periodic orbits can be labeled symbolically by strings of digits, each
between $0$ and $M-1$, the $T$-th digit indicating the horizontal strip
in which the particle is found at time $T$. A simple nontrivial
example is the $\ldots11111\ldots$ orbit for the case $M=3$. The
instability exponent of this orbit is given by $|\log(w_1)|$, and
homoclinic orbits of various instabilities and phases can be constructed
({\it e.g.} $\ldots111112001202011111\ldots$). This is in fact
the periodic orbit that
we will use for the numerical data on scarring in Section~\ref{genbakernum}.

A quantum mechanical version of this system can be obtained readily for
any value of Planck's constant given by $h=1/N$. $N$ is the dimension of the
resulting Hilbert space. The position basis consists of states $|j\rangle,
\,j=0\ldots N-1$, corresponding semiclassically to vertical strips located at
$x_j=(j+\epsilon_1)/N$. Similarly, the momentum
basis is formed by $|\tilde{k}\rangle$, $\tilde{k}=0\ldots N-1$,
with the states
living at $p_{\tilde{k}}=(\tilde{k}+\epsilon_2)/N$. The two bases are related
by a discrete fourier transform. The numbers $\epsilon_{1,2}\in[0,1)$
are constants which
provide the quantization conditions for the system (they define the
phases associated with going around the torus in the vertical and
horizontal directions, respectively).

To define the baker's map dynamics we write $N$ as a sum of integers
$N=\sum_{m=0}^{M-1} N_m$,
approximating the classical division of phase space into strips
($|N_m-w_mN|<1$). Then the leftmost $N_0$ position states are mapped
into the bottom $N_0$ momentum states by a discrete fourier transform, and
similarly for each of the other strips. Finally, we transform back to the
position basis. So the one-step evolution operator in the position basis
has the form
\begin{equation}
U=\left[ F^{-1}_N \right] \left[ \begin{array}{cccc}
e^{i\theta_0} F_{N_0} & 0 & \ldots & 0 \\
0 & e^{i\theta_1} F_{N_1} & \ldots & 0 \\
\vdots & \vdots  & \ddots & \vdots \\
0 & 0 & \ldots & e^{i\theta_{M-1}} F_{N_{M-1}} \end{array} \right] \,,
\end{equation}
where $F_N$ is a discrete fourier transform matrix on $N$ sites, and
the $\theta_m$ are arbitrary angles. Different choices of $\theta_m$
correspond to different semiclassical theories, all having the same
classical limit. The statistical properties of the quantum mechanical
system should be independent of the choice of  quantization parameters
$\epsilon_{1,2}$ and $\theta_m$, so these can be randomly chosen
in the context of an ensemble averaging.

The semiclassical one-step propagator can be easily written down
in the position basis in terms of the stretching factors $w_m$ and
phases $\theta_m$, making use of the
symbolic dynamics governing this system.
\begin{equation}
U_{\rm SC}(x',x)=\sum_m \sqrt{w_m} e^{i s_m x'/\hbar+ i\theta_m}
\delta(x-(s_m+w_m x'))\,.
\end{equation}
The exact semiclassical $T-$step propagator is obtained by iterating
the one-step formula $T$ times (this is permitted
as long as we do not impose the
quantization condition by forcing $x$ at intermediate times
to have one of the $N$ quantum
mechanically allowed values). Long-time overlaps of Gaussian
wavepackets, as well as semiclassical spectra and eigenstates can be
computed in an efficient manner (for example, by fourier transforming to
momentum variables and cutting off the high-frequency modes)\cite{expwall}.
These calculations generally show good agreement
with the quantum results, although the convergence is not uniform over
phase space~\cite{scspec}.
In particular, the semiclassical approximation does not
see the ``diffractive" effects associated with the boundaries between
classical strips.

To illustrate the effect of isolated returning orbits as discussed in
Section~\ref{isolrec}, we construct a modified version of the system described
above. A rectangle is divided into $2L+1$ strips numbered $-L\ldots +L$. The
strips $-L\ldots -1$ all have the same width, and similarly for strips numbered
$+1\ldots +L$. Now the three vertical strips $-L$, $0$, and $+L$ get mapped
into the horizontal region spanned by strips $-1$, $0$, and $+1$, via the
usual generalized baker's map dynamics for $M=3$ described
in the above paragraphs. Simultaneously
strip $i$ for $1<i<L-1$ gets shifted to the right
into strip $i+1$, and similarly
$-i$ is shifted to the left into $-(i+1)$.
In effect, we have two long corridors
to the left and right of the hyperbolic region, each of which must
be traversed in its entirety before one returns to the center, which is where
all of the mixing occurs. This slows
down the total mixing in phase space because each excursion away from strip $0$
takes at least $L$ steps. An effective symbolic dynamics using the three
symbols `$0$', `$-$', and `$+$' can be used to represent the trajectories,
where `$+$'$=+1\,+2\,\ldots\,+L$, and similarly for `$-$'. Then for the
periodic orbit $\ldots0000\ldots$, the shortest homoclinic excursions
involve $L$ steps, and are symbolically represented as
$\ldots000+000\ldots$ and $\ldots000-000\ldots$. On the other hand, the
Lyapunov exponent of the original orbit (and therefore the decay time
of the linear dynamics) can be arranged to be of order unity, creating
a clean separation of scales between the two effects. Numerical results
for this system will be presented towards the end of Section~\ref{genbakernum}.

\section{Numerical results}
\label{genbakernum}

We now proceed to examine the numerical evidence for scarring in
the three-strip generalized baker's map described in the previous section.
We concentrate on the period one orbit $\ldots11111\ldots$, for which the
particle stays always in the middle strip.
The location of the fixed point is given by
$x_{\rm FP}=p_{\rm FP}=w_0/(w_0+w_2)$, and the
Lyapunov exponent is $\lambda=|\log{w_1}|$. An ensemble average can
be performed over different strip widths $w_0$, $w_1$, and $w_2$, and
over the value of Planck's constant $\hbar=1/2\pi N$. The integers
$N_m$, describing the dimensions
of the subspaces corresponding to the three strips, are chosen
to be prime to eliminate possible sources of anomalous behavior.
(In the original baker's map, such anomalous behavior is associated with
values of $N$ which are divisible by powers of 2; see for example
Ref.~\cite{expwall}.)
Also, we demand $N_0\ne N_2$ to eliminate the parity symmetry.
This parity symmetry, also present in the original baker's map and given
by $P: x\to 1-x, \; p\to 1-p$, would otherwise produce a factor of $2$
enhancement in the inverse participation ratio for the $\ldots11111\ldots$
orbit, as explained in the concluding paragraph of Section~\ref{heissym}.

For each of $97$ realizations of this system, the inverse participation
ratio was computed for a circular wavepacket centered on the periodic point.
The Lyapunov exponent for the orbit ranged from a low value of $0.28$
(corresponding to a linear IPR enhancement factor of $10.4$ according
to Eq.~\ref{iprval}, and not including the quantum fluctuation factor $F$),
to a high value of $1.94$, corresponding to a linear prediction of
$1.66$ for the IPR. The values of $N$ in this ensemble lie between $129$
and $419$. In Figure~\ref{iprfig}, the actual quantum value of the IPR
is plotted (using squares) {\it versus}
the linear prediction on the horizontal axis.
The data is compared to a line
of slope $2.2$, corresponding to the quantum fluctuation factor $F$
appropriate to this system. $F$ is obtained here by measuring the mean IPR
for a wavepacket placed randomly in the phase space instead of on
a periodic point. We see
large fluctuations in this Figure around the predicted behavior,
but the overall linear trend is certainly correct. Fluctuations in the
amount of scarring at a given value of the Lyapunov exponent are of the order
of twenty percent around the mean. This level of variation in scarring
strength from orbit to orbit is quite reasonable taking into account
the fact that
the effective number of spectral lines under the linear envelope is well
under $100$ in most cases studied. (This effective dimension of the space in
which the wavepacket lives is given by the total dimension $N$
divided by the IPR of
the linear envelope.)

For a subset of $46$ out of this set of $97$ systems, the semiclassical IPR
was also computed (this is defined by the fluctuation of the {\it right}
eigenstates of the time evolution). These values are shown on the
same plot in Figure~\ref{iprfig} using the `+' symbol.
The semiclassical IPR is correlated with but does not exactly follow
the full quantum value. Deviations from the predictions of the random
statistical theory
are present in the semiclassical as well as in the quantum calculation. This
suggests that cases of excessive (or deficient) scarring must be associated
with non-randomness in the new long-time recurrences, and not simply with
a breakdown of the semiclassical approximation. Given the small number
of parameters defining the system, and the finite dimension $N$
of the Hilbert space, such occasional anomalous
behavior is not very surprising.

On the same Figure, we also plot (using triangles)
the results for a modified version of the
generic baker's map, where random matrices
have been substituted for the matrices $F_{N_0}$ and $F_{N_2}$
describing the dynamics of the left and right strips (while
leaving the behavior of the middle strip unchanged). This corresponds to
explicitly randomizing the long-time recurrences, while preserving the
constraint imposed by the short-time dynamics in the region of the fixed
point. The full IPR in this case follows the same linear dependence
on the short-time envelope prediction, but with a clearly smaller
numerical coefficient $F$. This is consistent with the finding that the
generic wavepacket-averaged (non-scarred) value of the IPR for this
system is smaller ($2.0$ compared with $2.2$ for the true baker's map).

In Figure~\ref{iprvsn}, the ratio between the actual IPR and the linearly
predicted value is plotted (squares) as a function of the effective
system size $N_{\rm eff}=N/{\rm IPR}_{\rm lin}$.
We find significant fluctuations around the mean value of $2.5$, with
no evident trend as $N_{\rm eff}$ increases. In a random matrix theory model,
fluctuations around the mean value of the IPR scale as
$1/\sqrt{N_{\rm eff}}$, and
this behavior is not inconsistent with the data,
although the range of $N$ used is
not sufficient to see the convergence. On the same Figure, the corresponding
ratio is plotted (using triangles)
for the ``randomized baker's map" described in the previous
paragraph. Here the fluctuations around the mean are only slightly
smaller than for
the real baker's map (their magnitude being
about $15$ rather than $20$ percent). The
mean value of the full-to-linear-IPR ratio is also smaller, as explained
in the previous paragraph. For the explicitly randomized system, the only
correlations in the new long-time recurrences (after having taken into
account the short-time constraint) are due to the finiteness of the
Hilbert space, and are expected to become insignificant in the
$\hbar \to \infty$ limit.

So far we have been focusing on the second moment of the spectral
intensity distribution, for various realizations of the generalized
baker's map system. We now turn to a specific system, for which we
will be able to look at the entire spectrum and see explicitly the
pattern of fluctuations around the linear envelope. We choose the first
entry in our data file, with $N=223$ and three strips of widths $N_0=79$,
$N_1=101$, and $N_2=43$. The Lyapunov exponent for the $\ldots11111\ldots$
orbit is $0.79$, producing by Eq.~\ref{iprval} a linear envelope with
an IPR of $3.97$. The actual quantum value of the IPR is $7.93$, so there is
a Heisenberg-time fluctuation factor of $2.0$ (compared to the
expected value $2.2$). The semiclassically
computed value for the IPR is somewhat larger, at $8.46$.

The intensity spectrum for a circular wavepacket centered on the
fixed point is plotted in Figure~\ref{specfig}a, along with the
linear envelope $S_{\rm lin}(E)$ and an intermediate envelope
obtained by fourier transforming $A(T)$ for $|T|<30$.
The intermediate envelope corresponds to a time scale large compared
to the linear decay time and mixing time of the system,
but still small compared with the
Heisenberg time.
A semiclassical
version of this intermediate envelope is also plotted and is seen to
be very similar to the full quantum computation.
In Figure~\ref{specfig},
a portion of the line spectrum is compared with the same spectrum
computed semiclassically. (Again, the semiclassical spectrum is defined
by taking the overlaps between the wavepacket and the right eigenstates
of the semiclassical evolution. The imaginary parts of the semiclassical
energies are ignored.) In Figure~\ref{specfig}c, the quantum spectrum
of Figure~\ref{specfig}a is now
plotted after dividing out by the linear envelope. We
see that the underlying oscillations around the envelope are in fact
energy independent, and are equally strong near the peak and valley of the
envelope.

In Figure~\ref{histo}, a histogram of these scaled intensities is
plotted, after averaging over several realizations of the system.
This is compared to a similar histogram of the unscaled intensities. It
is seen that the former comes much closer to obeying an exponential
law as predicted by Porter-Thomas.

In Figure~\ref{ecorr}, the spectral correlation function
$<S(\epsilon+E)S(\epsilon)>/<S(\epsilon)>^2$ is plotted in the form of
a histogram (plusses), and compared to the correlation function for a
scaled spectrum (diamonds). An ensemble average and an average over the
energy $\epsilon$ has been performed in each case. We can see that the
correlation function for the scaled spectrum is uniform with small random
fluctuations (we ignore the large correlations for $E$ of the order of
a mean level spacing, $E=O(1/N$)). On the other hand, correlations in the
unscaled spectrum are very striking, and sharply peaked near $E=0$.

In Figure~\ref{figisol}, a spectrum with a linear and a second-order envelope
is plotted for a system with isolated homoclinic recurrences, associated
with excursions away from the mixing region
as discussed at the end Section~\ref{genbak}. We choose a minimum excursion
length $L=23$, with strip $0$ having width $N_0=17$ and all the others
being of width $N_{\pm 1}=\ldots=N_{\pm 23}=19$. A gaussian wavepacket
is launched at the center of the middle strip. The initial linear decay
occurs within $1$ -- $2$ time steps
(the Lyapunov exponent at the fixed point being
$1.17$), but the initial recurrence, corresponding to the homoclinic orbit
$\ldots 0000+0000 \ldots$ and its parity counterpart $\ldots 0000-0000 \ldots$
does not peak until $T=28$. The next set of recurrences results from the
classical orbits $\ldots 0000+0^n+0000 \ldots$ and
$\ldots 0000+0^n-0000 \ldots$, and their parity counterparts ($n$ being a
nonnegative integer). In all, $12$ identifiable peaks with spacing $\sim 23$
can be identified, the structure not breaking down completely until
$T \sim 300$, at which time random mixing begins to take over (classically,
homoclinic excursions with more and more insertions of the `$0$' symbol
are becoming important here for entropic reasons). The squared
autocorrelation function for the wavepacket
is plotted in Figure~\ref{figisol}.

\begin{figure}
\vskip .2in
\vskip .2in
\caption{The spectrum corresponding to the wavepacket whose return probability
is plotted in the previous Figure, along with a linear envelope (dotted line),
an intermediate envelope corresponding to the first two sets
of homoclinic
recurrences (dashed line), and higher resolution envelope
corresponding to the first
six sets of recurrences (solid line).}
\label{specisol}
\end{figure}
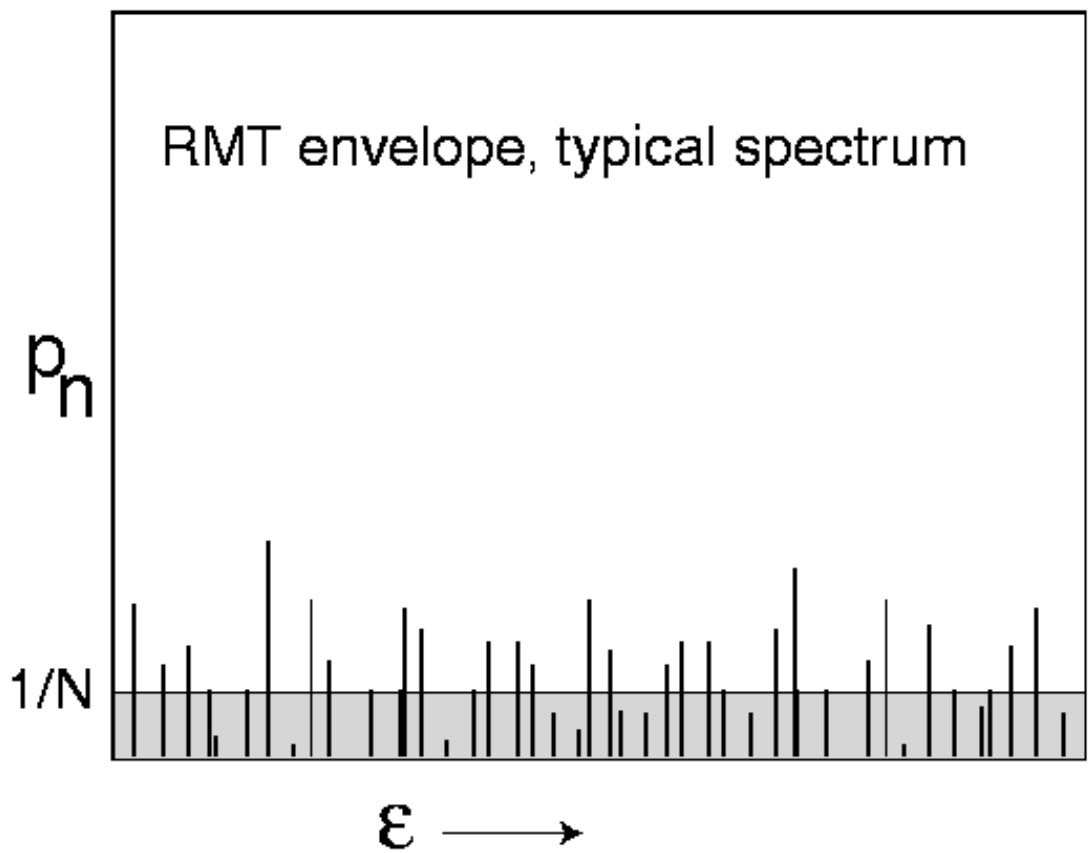
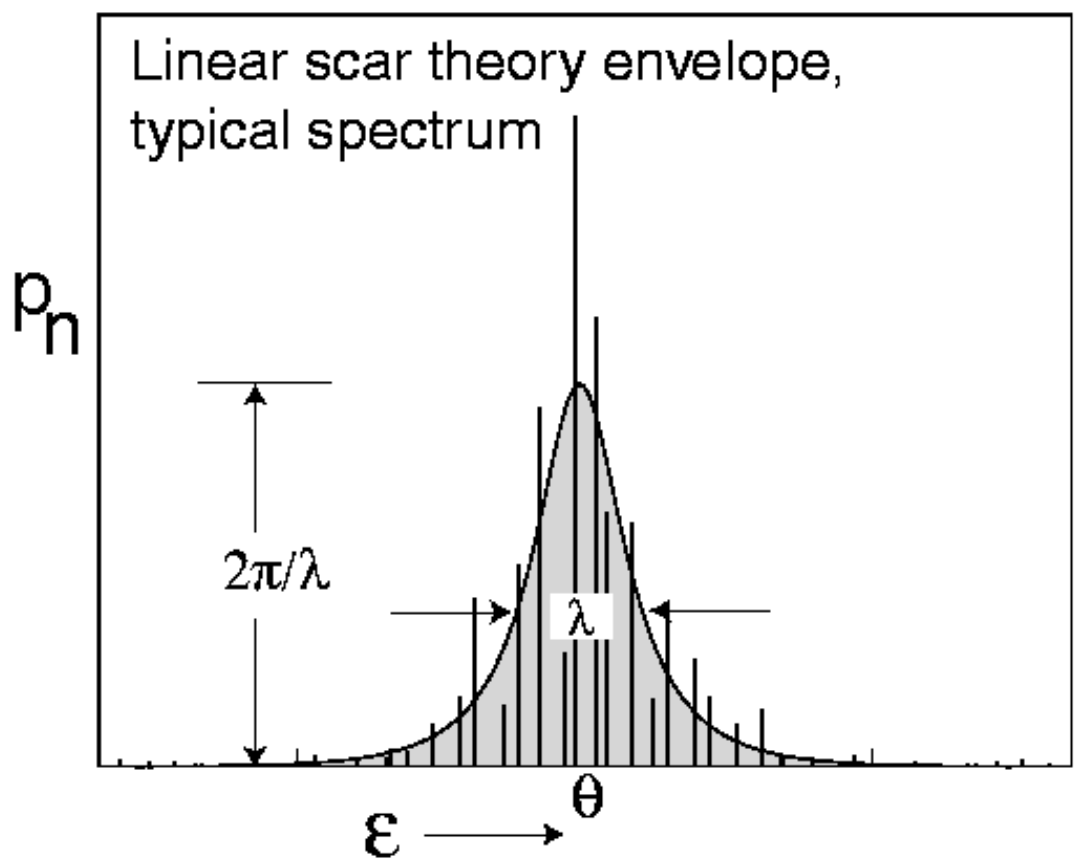
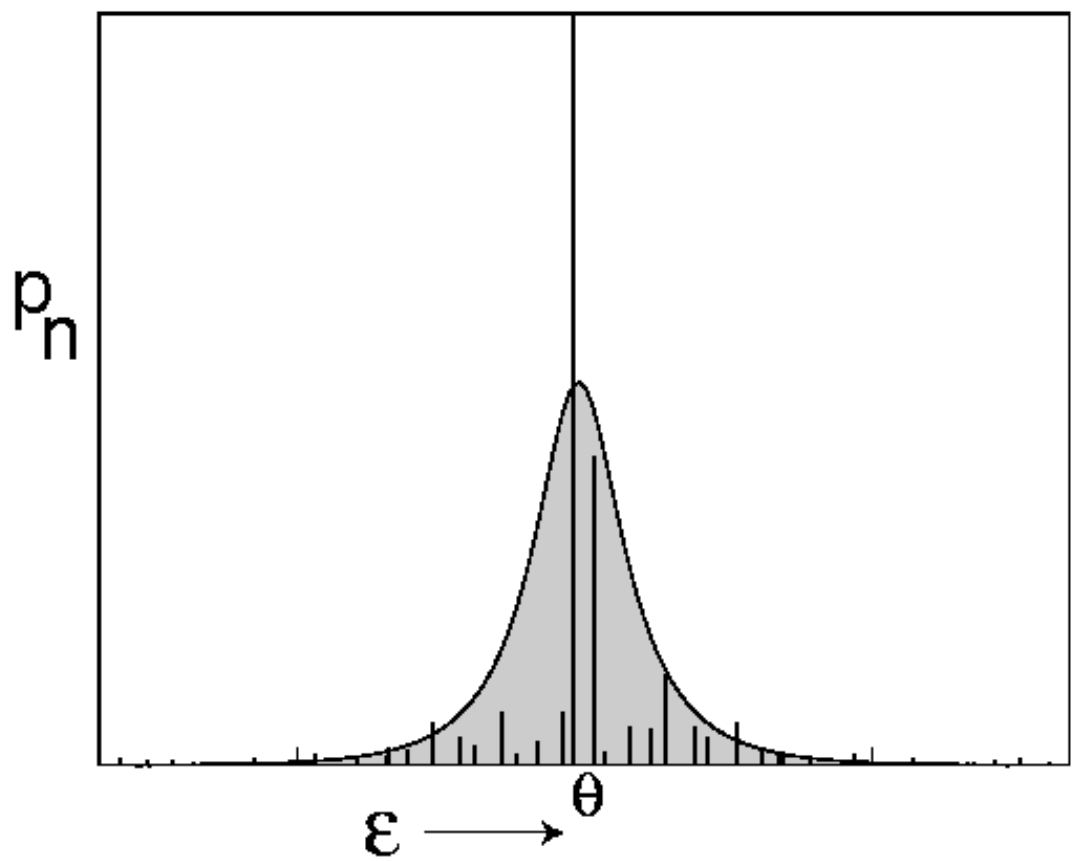
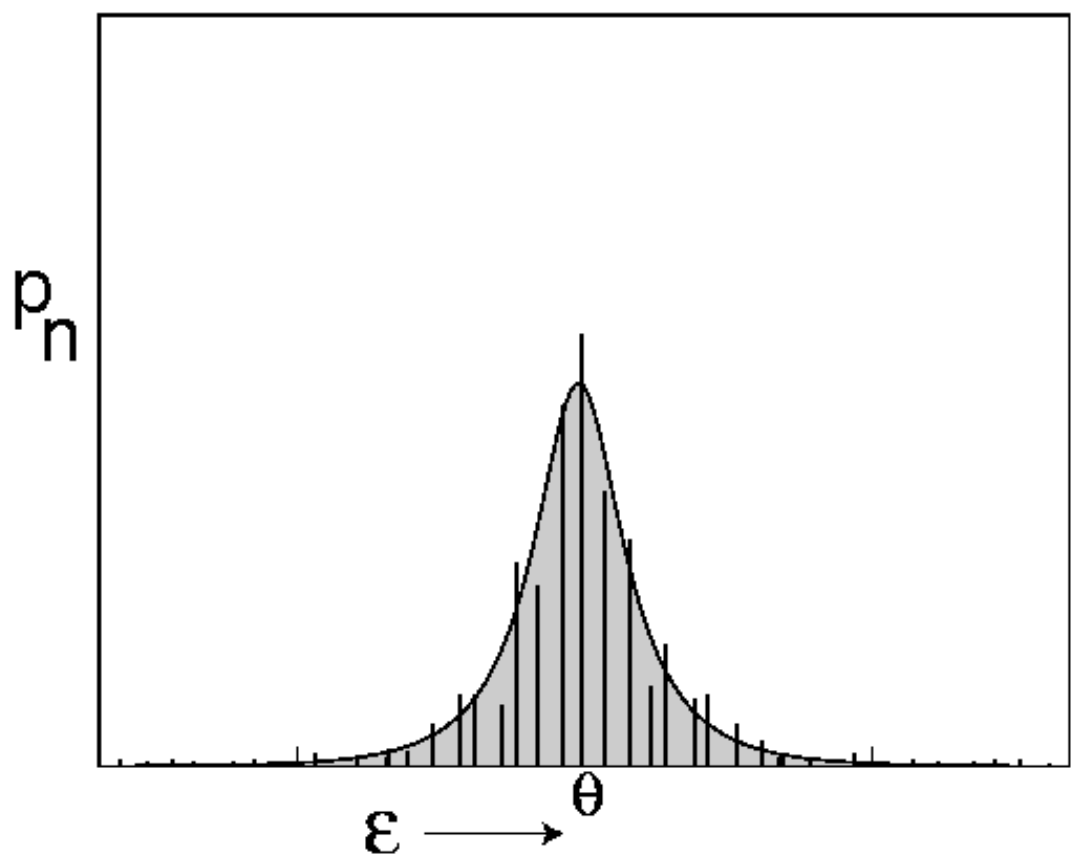
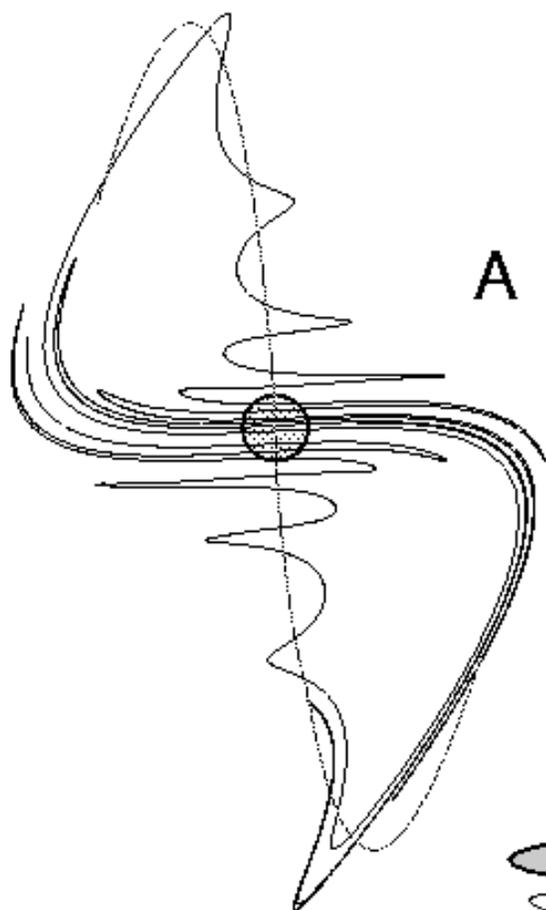
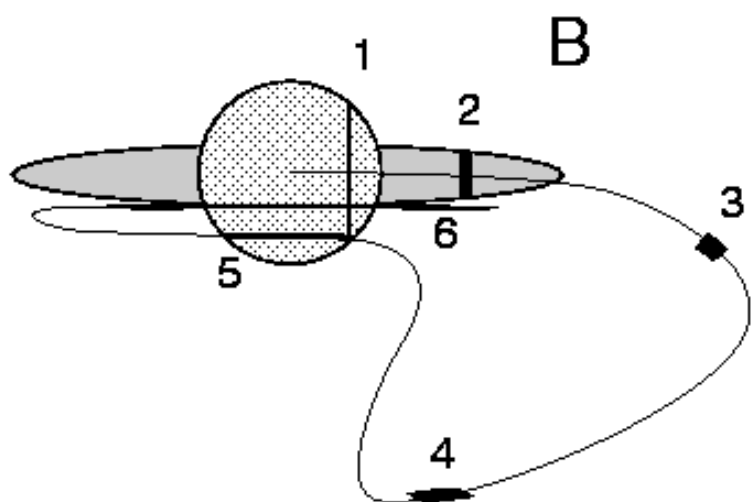
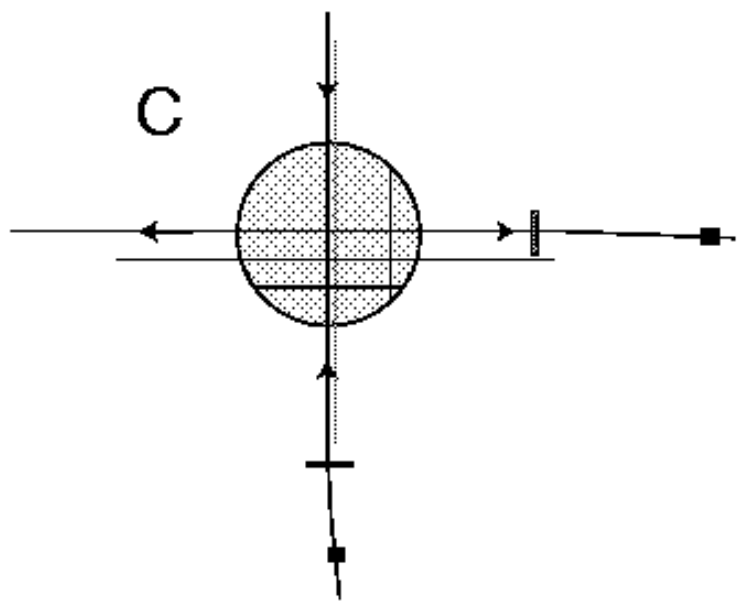
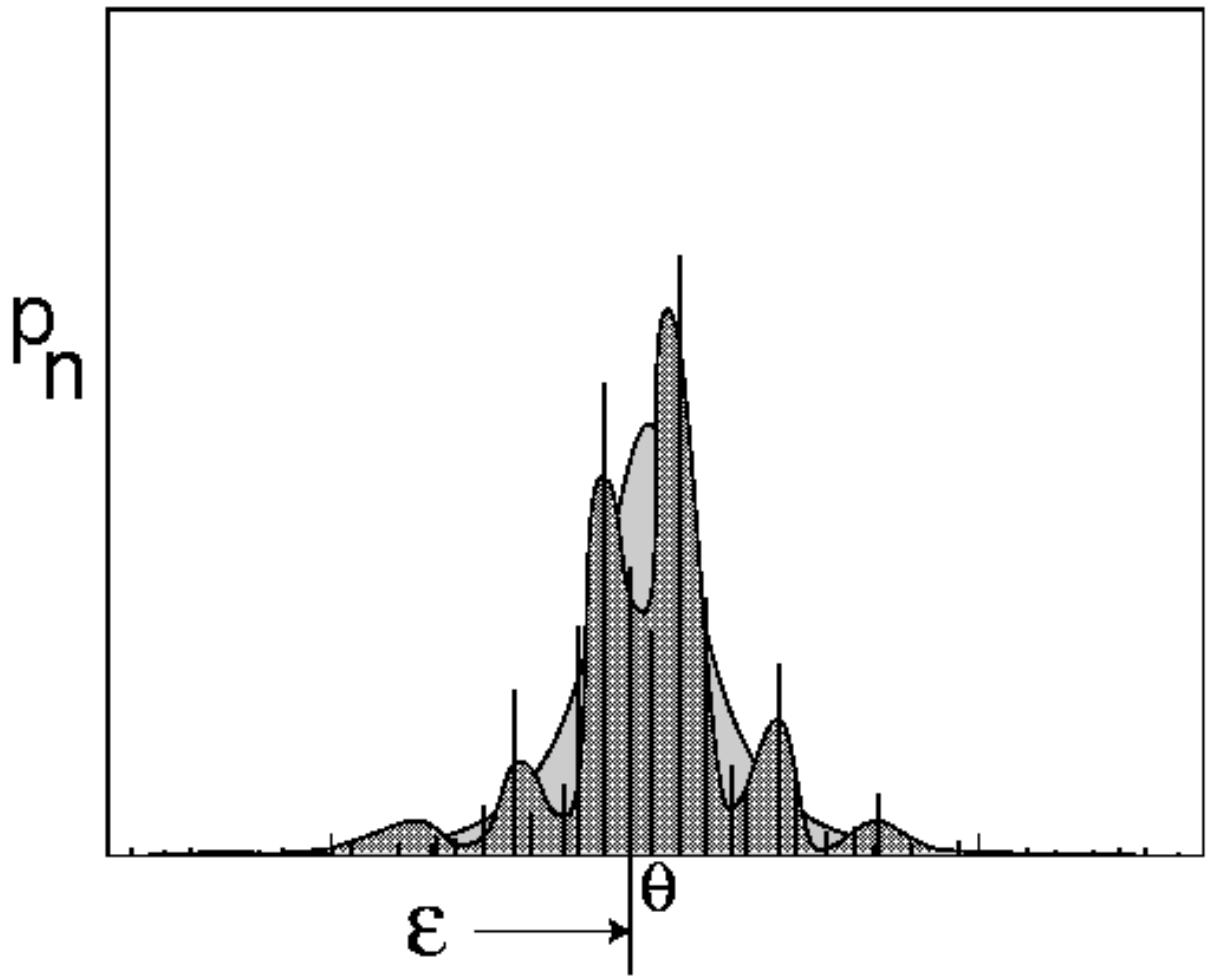
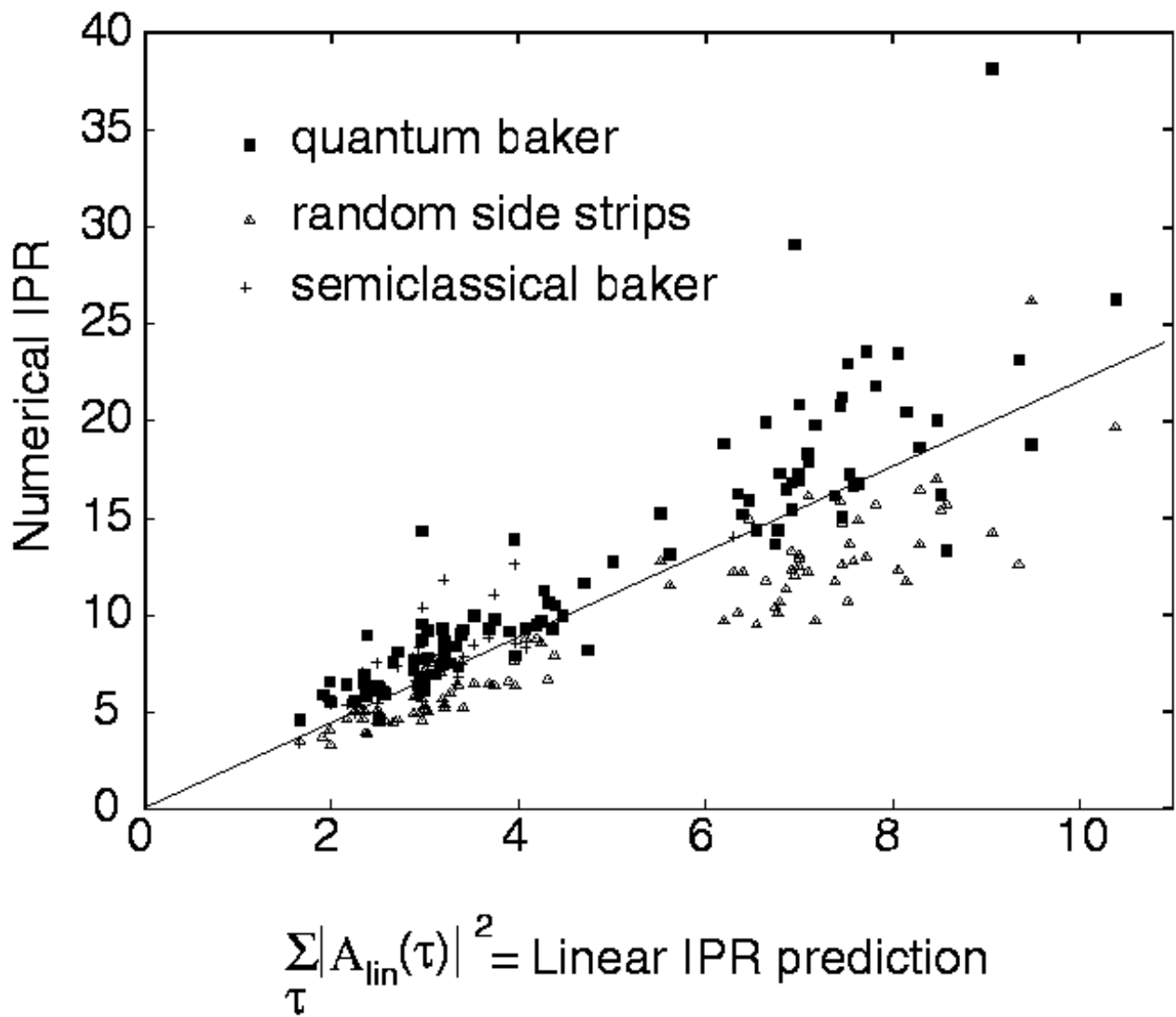
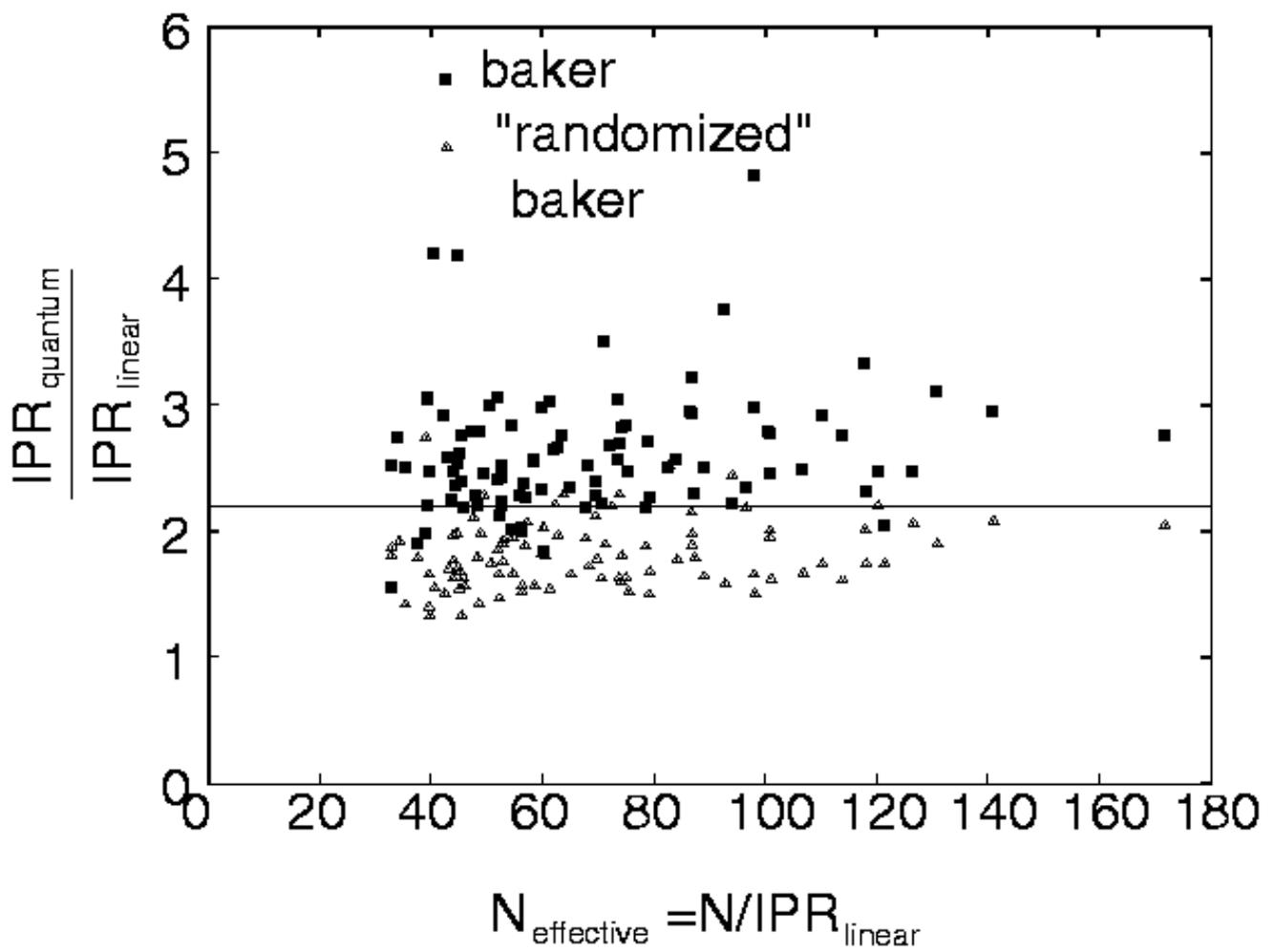
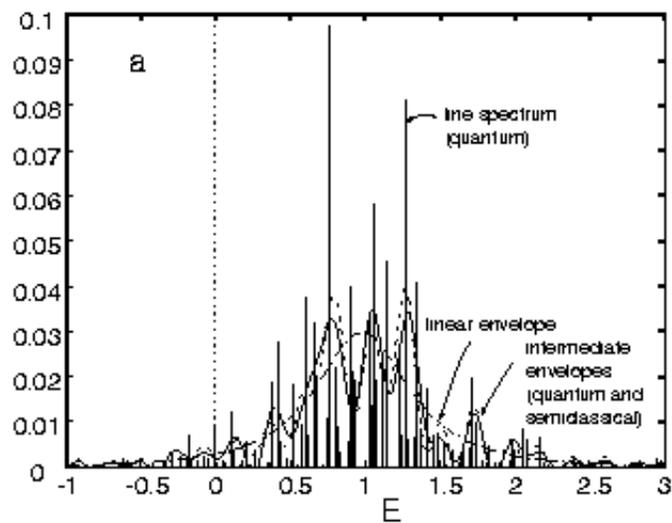
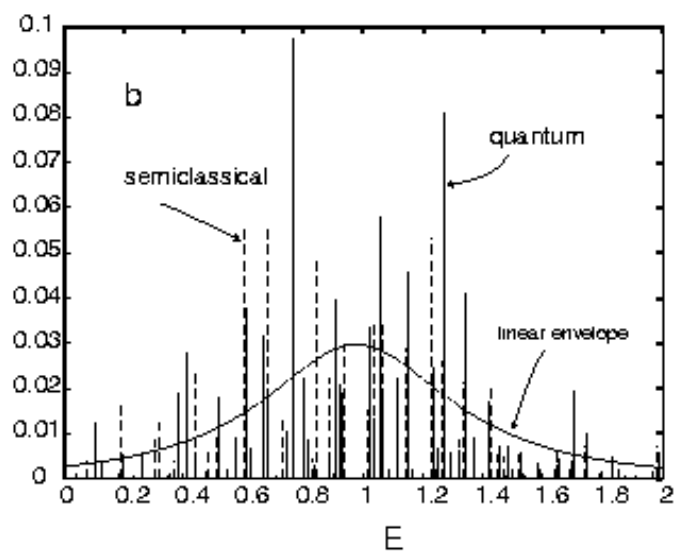
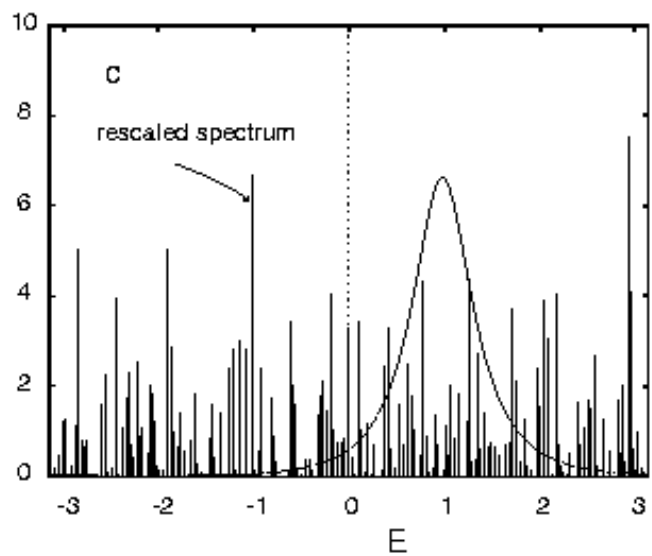
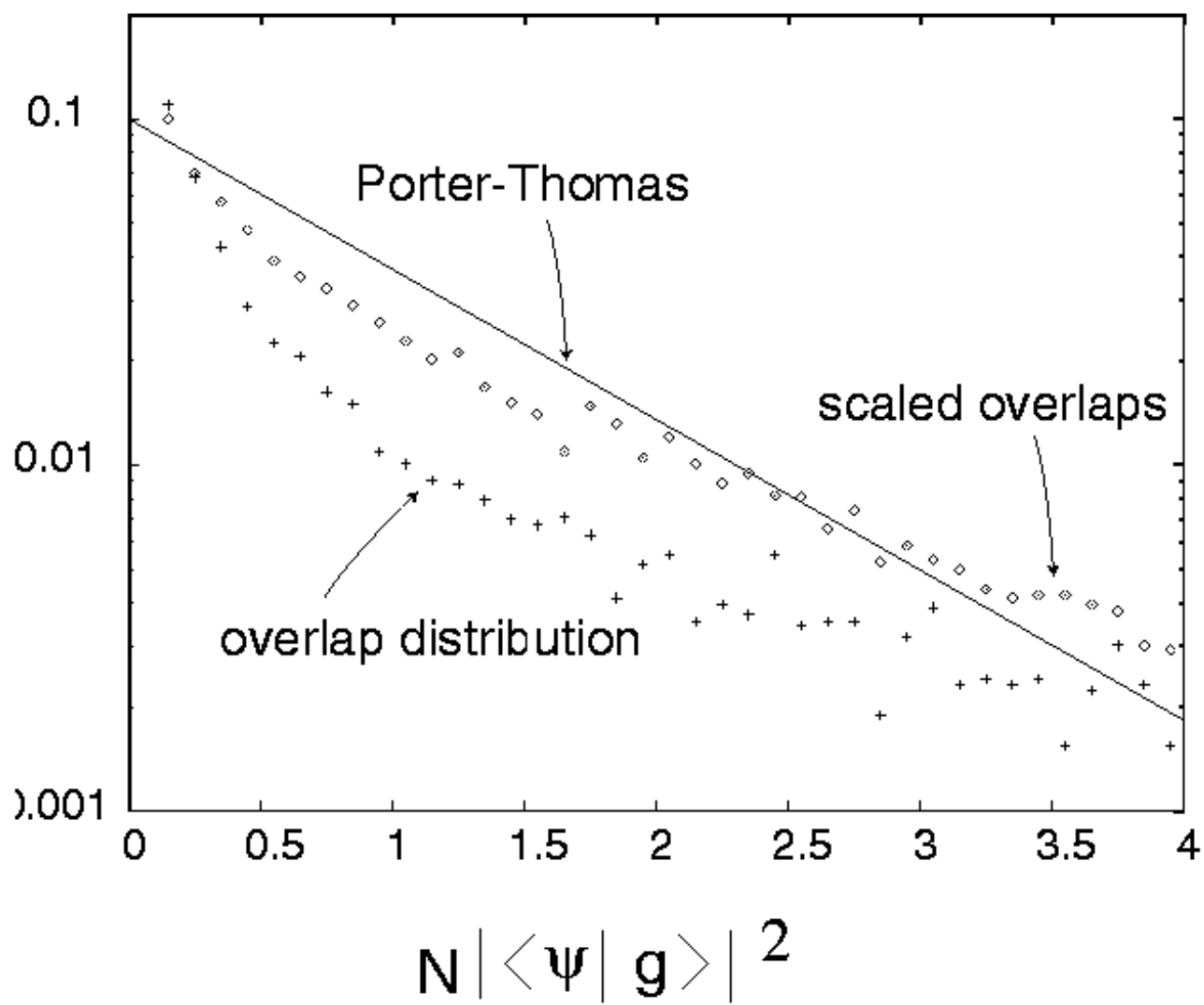
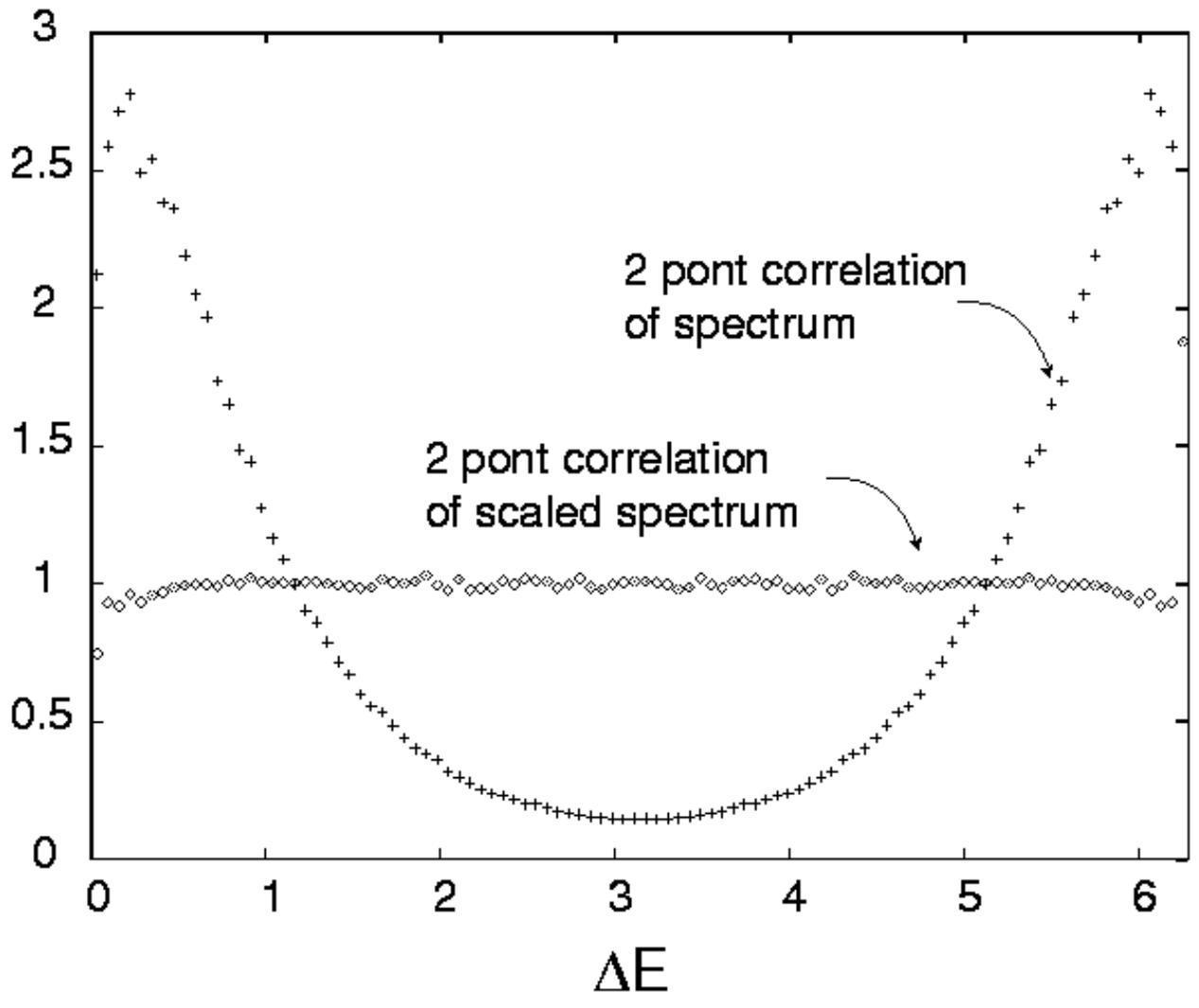
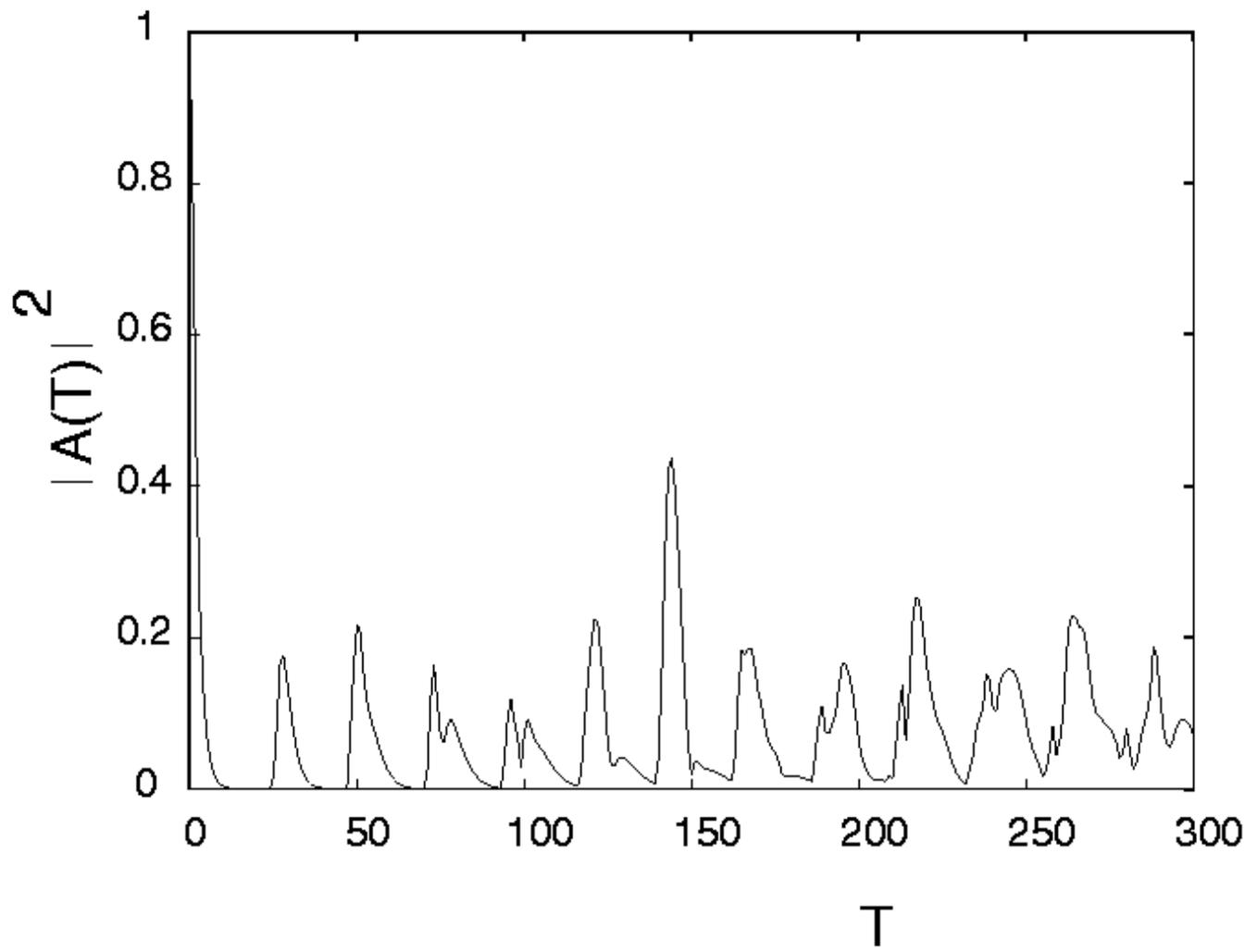
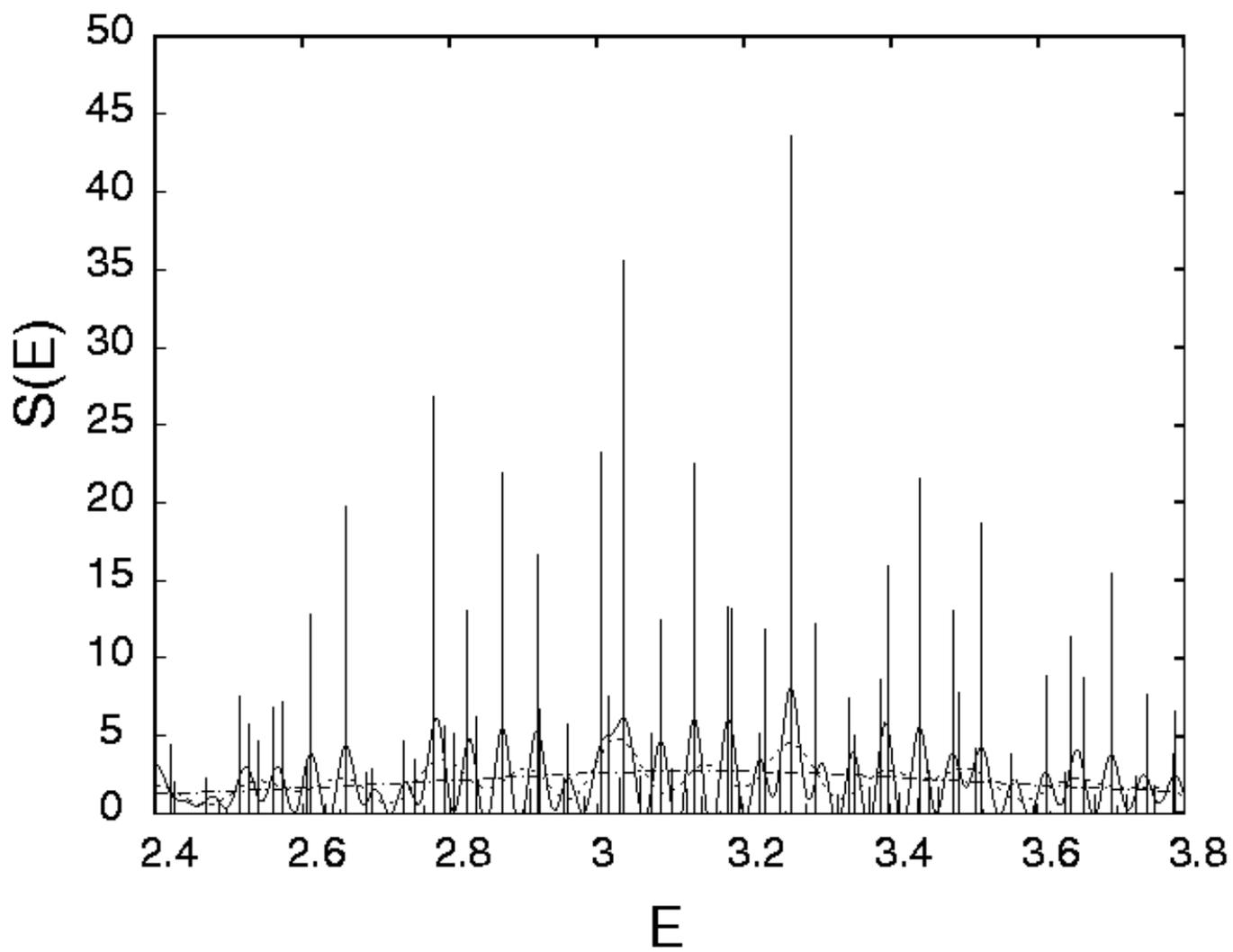

In Figure~\ref{specisol}, the corresponding energy spectrum is plotted
along with a linear envelope (dotted line) and an envelope resulting
from the first two sets of recurrences (up to two insertions of `$+$'
or `$-$'), dashed line. The solid curve includes the effects of the
first six sets of homoclinic returns.

\section{Conclusion}

By focusing on the importance of nonlinear recurrences, long-time
fluctuations, symmetry factors, and the local classical structure
around a periodic orbit, we have attempted to clear up some of the
long-standing mysteries in the literature on scarring. We have seen
that the linear theory, even in a worst-case ``egalitarian" scenario,
makes a lower bound prediction on scarring strength that is a function
of the instability of the orbit only, and independent of energy and $\hbar$.
Generically, we expect random long-time fluctuations to be present,
associated with nonlinear excursions away from the periodic orbit.
When these gaussian random fluctuations are included in the theory,
quantitative agreement
between theory and numerics is obtained using
measures such as the inverse participation ratio, wavefunction intensity
distribution, and correlations in the local density of states.
Scarring stronger than that predicted by the random nonlinear theory
can be obtained in the presence of identifiable homoclinic
recurrences.

The formalism developed here lends itself naturally to the investigation
of other effects of short-time dynamics on the properties of quantum
eigenstates. Such short-time behavior may involve classical structures
other than periodic orbits, diffraction, and ``quiet time" behavior.
Intensity correlations among eigenstates as well as phase space
correlations in the structure of individual eigenstates can be studied.
Interesting questions arise in the design of ``optimal" measures for
observing this class of deviations from random matrix theory
behavior.

The nonlinear scarring theory as presented in this paper has been
used recently to study the tail of the wavefunction intensity distribution
in chaotic systems\cite{intdistr}.
Predictions, validated by numerical experiments,
have been obtained for the distribution of eigenstate intensities in a single
region of phase space, for phase space-averaged distributions, and also
for ensembles which include systems with orbits of different lengths
and instability exponents. Power-law tails are naturally obtained in
the process of ensemble averaging, under certain assumptions.

\section{Acknowledgments}

This research was supported by the National Science Foundation under
grant number CHE-9321260. We wish to thank the Institute
for Theoretical Physics at UCSB, where this research was initiated,  for its
hospitality. We also wish to thank the Isaac Newton Institute for the
Mathematical
Sciences in Cambridge, where this work was completed. EJH would like to thank
S. Tomsovic for many stimulating conversations.

\begin{figure}
\vskip .2in
\vskip .2in
\caption{The short time dynamics of the localized wavepacket
imposes an envelope in the
local density of states which
the resolved spectrum (coming from long-time dynamics up to times of order
of the Heisenberg time) must obey.  The envelope has a peak at quasi-energy
$\epsilon=\theta $, a width $\delta \epsilon\ \sim{\cal O}
(\lambda)$, and a height $\sim {\cal O}  \ (1/\lambda)$.
 }
\label{envel}
\end{figure}

\begin{figure}
\vskip .2in
\vskip .2in
\caption{The short time dynamics by itself does not predict whether
``totalitarian'' (top) or ``egalitarian'' (bottom) filling of the
local density of states envelope occurs. Both spectra have the same
short time local density of states envelope.}
\label{envel2}
\end{figure}
\begin{figure}
\vskip .2in
\vskip .2in
\caption{ Nonlinear recurrences resulting from following
a homoclinic classical
orbit. Linear wavepacket spreading is followed by a nonlinear excursion, and
finally by an approach back to the fixed point along the stable manifold.
Three levels of abstraction are shown. Diagram A  shows a portion of the
stable manifold, and a longer portion of the unstable manifold.  B is
more schematic, showing how a small rectangle in the initial disk
(representing the initial state in phase space) is compressed and
stretched, finally returning along the stable manifold.  Note that
this particular rectangle is special in that it returns soon,
guided by a primary homoclinic trajectory.   In C, we see a
regularized (normal form) version of B.  Two types of correlation
are seen in this Figure.  In B, for example, ``1''  and ``5'' have
exactly the same phase relation as ``2'' and ``6''.  Also ``5''  and
``6'' are related by a
phase associated with one iteration of the periodic orbit.}
\label{mani}
\end{figure}
\begin{figure}
\vskip .2in
\vskip .2in
\caption{A strong, isolated recurrence at a time later than the
Lyapunov decay time causes additional structure in the
spectrum at a higher resolution.  This new, oscillating envelope
 further constrains the possible behavior of the resolved spectrum.
 The final spectrum   has structure on at
least three scales.}
\label{nonlinfig}
\end{figure}
\begin{figure}
\vskip .2in
\vskip .2in
\caption
{A plot of the actual value of the inverse participation
ratio (IPR) for a wavepacket centered on a periodic orbit (squares),
{\it versus}
the value predicted by the linear theory. IPR's in the semiclassical
approximation are plotted using the `+' symbol, and IPR's in a baker's
map with random matrix theory nonlinear behavior are plotted with triangles.}
\label{iprfig}
\end{figure}
\begin{figure}
\vskip .2in
\vskip .2in
\caption{Here the ratio of the full IPR to the value predicted
by the linear theory is plotted against the effective number of states
that the wavepacket can overlap with (according to the linear theory).
Using triangles, the same quantity is plotted for a randomized baker's map.}
\label{iprvsn}
\end{figure}
\begin{figure}
\vskip .2in
\vskip .2in
\caption{In (a), the full spectrum is plotted along with the
linear envelope (dotted line), an intermediate envelope corresponding to
$|T|<30$ (solid line), and a semiclassical intermediate envelope (dashed
line). In (b), a portion of the spectrum (solid) is compared to the
spectrum obtained using semiclassical eigenstates (dashed). In (c),
the quantum spectrum has been divided out by the linear envelope of (a).}
\label{specfig}
\end{figure}
\begin{figure}
\vskip .2in
\vskip .2in
\caption{A histogram of scaled spectral intensities (diamonds)
after having divided out by the linear envelope, compared to a histogram
of raw (unscaled) intensities (plusses). The Porter-Thomas exponential
law is plotted as a solid line. All distributions are defined to have a mean
value of one.}
\label{histo}
\end{figure}
\begin{figure}
\vskip .2in
\vskip .2in
\caption{The two-point spectral correlation function of the
scaled spectrum (diamonds) is compared to the correlation function of the
raw spectrum (plusses), after ensemble and energy averaging.}
\label{ecorr}
\end{figure}
\begin{figure}
\vskip .2in
\vskip .2in
\caption{The quantum return probability as a function of time
for the modified baker's map with a minimum excursion length of $L=23$.}
\label{figisol}
\end{figure}
\end{document}